\documentclass[submission,Phys]{SciPost}

\usepackage{placeins} 

\usepackage[utf8]{inputenc} 
\usepackage[T1]{fontenc} 	
\usepackage[english]{babel} 

\usepackage[bitstream-charter]{mathdesign}
\usepackage{geometry} 		
\usepackage{amsmath} 		
\usepackage{mathtools} 		
\usepackage{float} 			
\usepackage{graphicx} 		
\usepackage{tabularx} 		
\usepackage{booktabs} 		
\usepackage{color, xcolor} 	
\usepackage{pdfpages} 		
\usepackage{extarrows} 		
\usepackage{multirow} 		
\usepackage{multicol} 		
\usepackage[subrefformat=parens]{subcaption}
\usepackage{enumitem} 		
\usepackage{xspace} 		
\usepackage{stackrel} 		
\usepackage{tikz} 			
\usetikzlibrary{positioning,calc}
\usepackage{braket} 		
\usepackage{bm} 			
\usepackage{tensor} 		
\usepackage{slashed} 		
\usepackage{siunitx} 
\usepackage{lastpage} 		
\usepackage{cite} 			
\usepackage[normalem]{ulem} 
\usepackage{fontawesome} 	
\usepackage{tocloft} 		
\usepackage{titlesec} 		
\usepackage{doi} 			
\usepackage{hyperref} 		
\usepackage[most]{tcolorbox} 					
\usepackage[nameinlink, capitalize]{cleveref} 	
\usepackage[nottoc, notlot, notlof]{tocbibind} 	
\usepackage[ruled, vlined]{algorithm2e} 		
\usepackage{makecell}
\usepackage{setspace}

\makeatletter
\def\BState{\State\hskip-\ALG@thistlm}
\makeatother

\makeatletter
\@ifundefined{pdfoutput}{}{\DeclareGraphicsRule{*}{mps}{*}{}}
\makeatother

\makeatletter
\DeclareRobustCommand*{\bfseries}{%
   \not@math@alphabet\bfseries\mathbf
   \fontseries\bfdefault\selectfont
   \boldmath
}
\makeatother

\hypersetup{
	pdftitle={Extrapolating Jet Radiation with Autoregressive Transformers},
	colorlinks=true, 			
	linkcolor={red!50!black}, 	
	citecolor={blue!50!black}, 	
	urlcolor={blue!80!black} 	
} 

\DeclareSymbolFont{usualmathcal}{OMS}{cmsy}{m}{n}
\DeclareSymbolFontAlphabet{\mathcal}{usualmathcal}

\SetArgSty{textnormal}
\SetKwComment{Comment}{{\small\#}~}{}
\SetCommentSty{mycommfont}

\setitemize{itemsep=0pt, parsep=0pt} 				
\setenumerate{itemsep=0pt, parsep=0pt} 				
\setitemize{itemsep=2pt,topsep=2pt,parsep=0pt,partopsep=0pt,leftmargin=*}
\setenumerate{itemsep=0pt,topsep=2pt,parsep=0pt,partopsep=0pt,labelindent=3pt,leftmargin=*}
\newcommand{\XLangle}{\Bigl\langle}
\newcommand{\XRangle}{\Bigr\rangle}
\newcommand{\XXLangle}{\biggl\langle}
\newcommand{\XXRangle}{\biggr\rangle}

\newcommand{\normal}{\mathcal{N}}

\newcommand{\loss}{\mathcal{L}} 	

\newcommand{\pmdj}[1]{p_{\theta,#1}}

\newcommand{\pd}{p_\text{data}}
\newcommand{\pkin}{\ensuremath{p_\text{kin}}\xspace}
\newcommand{\pbin}{\ensuremath{p_\text{bin}}\xspace}

\newcommand{\psplit}{\ensuremath{p_\text{split}}\xspace}
\newcommand{\pmd}{p_{\theta}}

\newcommand{\arXiv}[2][]{%
	\ifthenelse{\equal{#1}{}}%
	{\href{http://arxiv.org/abs/#2}{arXiv:#2}}%
	{\href{http://arxiv.org/abs/#2}{arXiv:#2~[#1]}}}

\newcommand{\gev}{\text{GeV}}

\def\slashchar#1{\setbox0=\hbox{$#1$}           
   \dimen0=\wd0                                 
   \setbox1=\hbox{/} \dimen1=\wd1               
   \ifdim\dimen0>\dimen1                        
      \rlap{\hbox to \dimen0{\hfil/\hfil}}      
      #1                                        
   \else                                        
      \rlap{\hbox to \dimen1{\hfil$#1$\hfil}}   
      /                                         
   \fi}

\newcommand{\tikznode}[2]{%
\ifmmode%
\tikz[remember picture,baseline=(#1.base),inner sep=0pt] \node (#1) {$#2$};%
\else
\tikz[remember picture,baseline=(#1.base),inner sep=0pt] \node (#1) {#2};%
\fi}

\def\mathswitchr#1{\relax\ifmmode{\text{#1}}\else$\text{#1}$\xspace\fi}
\def\mathswitch#1{\relax\ifmmode#1\else$#1$\xspace\fi}

\usetikzlibrary{arrows, shapes.geometric, arrows.meta, shapes, decorations.pathreplacing, fit, patterns, patterns.meta}

\definecolor{Rcolor}{HTML}{E99595}
\definecolor{Gcolor}{HTML}{C5E0B4}
\definecolor{Bcolor}{HTML}{9DC3E6}
\definecolor{Ycolor}{HTML}{FFE699}
\definecolor{Vcolor}{HTML}{CFB4E0}

\tikzstyle{expr} = [circle, minimum width=1.8cm, minimum height=1.8cm, text centered, align=center, inner sep=0, draw,font=\LARGE]
\tikzstyle{txt_huge} = [align=center, font=\Huge, scale=2]
\tikzstyle{txt} = [align=center, font=\LARGE]
\tikzstyle{cinn} = [double arrow, double arrow head extend=0cm, double arrow tip angle=130, shape border rotate=90, inner sep=0, align=center, minimum width=2.1cm, minimum height=2.3cm, fill=Bcolor, draw,font=\LARGE]
\tikzstyle{cinn_black} = [cinn, minimum height=2.5cm, fill=black]
\tikzstyle{arrow} = [thick,-{Latex[scale=1.0]}, line width=0.2mm, color=black]
\tikzstyle{line} = [thick, line width=0.2mm, color=black]
\tikzstyle{loss} = [rectangle, align=center,  minimum width=1.8cm, minimum height=1.5cm,fill=Rcolor,font=\LARGE, rounded corners]
\tikzstyle{xt} = [rectangle, align=center,  minimum width=4cm, minimum height=1.5cm,fill=Gcolor,font=\Large, rounded corners]
\tikzstyle{xts} = [rectangle, align=center,  minimum width=1cm, minimum height=1.5cm,fill=Gcolor,font=\Large, rounded corners]

\usetikzlibrary{arrows, shapes.geometric, arrows.meta, shapes, decorations.pathreplacing, fit}
\graphicspath{{./figs/}}
\begin{document}

\vspace*{-2.5em}
\hfill{}
\vspace*{0.5em}

\begin{center}{\Large \textbf{
Extrapolating Jet Radiation with Autoregressive Transformers
}}\end{center}

\begin{center}
Anja Butter\textsuperscript{1,2},
François Charton\textsuperscript{3},
Javier Mariño Villadamigo\textsuperscript{1},\\
Ayodele Ore\textsuperscript{1},
Tilman Plehn\textsuperscript{1,4},
and Jonas Spinner\textsuperscript{1}
\end{center}

\begin{center}
{\bf 1} Institut für Theoretische Physik, Universität Heidelberg, Germany \\
{\bf 2} LPNHE, Sorbonne Université, Université Paris Cité, CNRS/IN2P3, Paris, France \\
{\bf 3} Meta FAIR, CERMICS - Ecole des Ponts \\
{\bf 4} Interdisciplinary Center for Scientific Computing (IWR), Universität Heidelberg, Germany
\end{center}

\begin{center}
\today
\end{center}

\vspace{-1cm}
\section*{Abstract}
{\bf Generative networks are an exciting tool for fast LHC event
  fixed number of particles. Autoregressive transformers allow us to
  generate events containing variable numbers of particles, very much in
  line with the physics of QCD jet radiation, and offer the possibility to generalize to higher multiplicities. We show how transformers can
  learn a factorized likelihood for jet radiation and extrapolate in
  terms of the number of generated jets. For this extrapolation,
  bootstrapping training data and training with modifications of the likelihood
  loss can be used.}

\vspace{10pt}
\noindent\rule{\textwidth}{1pt}
\tableofcontents\thispagestyle{fancy}
\noindent\rule{\textwidth}{1pt}

\clearpage
\section{Introduction}
\label{sec:intro}

Modern LHC physics is defined by precision tests of the fundamental
properties of particles and their interactions, both in and beyond the
current Standard Model. The level of precision is continuously
improved by experimental and theoretical progress, accompanied by but
not limited to the rapidly increased LHC luminosity towards the
high-luminosity LHC.

In view of this precision program, experimental and theoretical LHC
physics are being transformed through modern machine learning
(ML)~\cite{Butter:2022rso,Plehn:2022ftl}. On the theory and simulation
side, a range of neural network applications are improving every step
of our first-principles simulation chain.  This includes phase-space
sampling~\cite{
  Bothmann:2020ywa,Gao:2020vdv,Gao:2020zvv,Danziger:2021eeg,Heimel:2022wyj,
  Bothmann:2023siu,Heimel:2023ngj,Deutschmann:2024lml,Heimel:2024wph},
  scattering amplitude surrogates~\cite{Bishara:2019iwh,Badger:2020uow,Aylett-Bullock:2021hmo,Maitre:2021uaa,Winterhalder:2021ngy,Badger:2022hwf,Maitre:2023dqz,Spinner:2024hjm,Maitre:2024hzp,Brehmer:2024yqw,Breso:2024jlt}, end-to-end event generation~\cite{
  Hashemi:2019fkn,DiSipio:2019imz,
  Butter:2019cae,Alanazi:2020klf,Butter:2023fov}, 
and detector
simulators trained on full simulations~\cite{Paganini:2017hrr,
  Paganini:2017dwg,
  Erdmann:2018jxd,Belayneh:2019vyx,Buhmann:2020pmy,
  Krause:2021ilc,
  ATLAS:2021pzo,Krause:2021wez,Buhmann:2021caf,Chen:2021gdz,
  Mikuni:2022xry,
  Cresswell:2022tof,Diefenbacher:2023vsw,
  Hashemi:2023ruu,
  Xu:2023xdc,
  Buhmann:2023bwk,Buckley:2023daw,Diefenbacher:2023flw,Ernst:2023qvn,Favaro:2024rle,Buss:2024orz,Quetant:2024ftg,Krause:2024avx}.

The main workhorses behind this transformation are generative
networks. They learn phase space densities or Jacobians from simple
distributions given sets of events and can reproduce these densities
through fast sampling. Modern generative network architectures are
normalizing flows, diffusion networks, and autoregressive transformers. Because they are fast, differentiable, and flexible, generative networks can enable new simulation and analysis strategies. These networks are reaching new levels of accuracy and can significantly amplify
simulated training data~\cite{Butter:2020qhk,Bieringer:2022cbs} and
speed up the generation. Given the LHC requirements, they have to be
controlled and precise in encoding kinematic patterns over an,
essentially, interpretable phase space~\cite{Butter:2021csz,
  Winterhalder:2021ave,Nachman:2023clf,Leigh:2023zle,Das:2023ktd}.
Conditional versions of the forward-generative networks allow for probabilistic
unfolding~\cite{Bellagente:2019uyp,
  Bellagente:2020piv,Backes:2022vmn,Leigh:2022lpn,Raine:2023fko,Shmakov:2023kjj,Diefenbacher:2023wec,Huetsch:2024quz}
or inference through posterior
sampling~\cite{Bieringer:2020tnw,Butter:2022vkj,Heimel:2023mvw}.

In this paper we tackle the physics problem of using generative
networks to describe jets radiated from a hard scattering process. In
fundamental QCD, jet radiation is described by successive
probabilistic parton splittings. It is an integral part of QCD
predictions for hadron colliders, where final states with a fixed
number of jets are not in line with parton densities and collinear
factorization~\cite{Ellis:1996mzs,Plehn:2009nd,Campbell:2017hsr}. The
corresponding splitting kernels and the generated phase space
correlations are approximately
universal~\cite{vanBeekveld:2022ukn}. The generated number of jets
follows well-defined patterns, also predicted by QCD.

Autoregressive generative networks can, just like with language,
generate open-end sequences of particles, or events with a variable
number of particles. An autoregressive generation requires a
factorized phase space probability\cite{Andreassen:2018apy,
  Andreassen:2019txo}. This structure matches the QCD aspects of
universal splittings and well-defined jet numbers. Our generative
architecture of choice is an autoregressive
transformer~\cite{Finke:2023veq,Butter:2023fov}. An attractive benefit of this approach is the possibility of exploiting universal structures across jet multiplicities, which could allow for a single generative network to be deployed instead of a collection of specialized models. In this work we establish the fundamental idea by studying extrapolation to higher multiplicities, highlighting both the challenges and the opportunities.

The goal of this paper is to show, for the first time, that a
generative transformer can extrapolate in the number of jets
and generate approximately universal jet radiation for higher jet
numbers than seen during the training. In Sec.~\ref{sec:basics} we
describe the QCD structures motivating an approximately factorized
phase space likelihood and its ML-realization, leading to our
autoregressive generative transformer. We then present extrapolated
predictions in the number of jets in Sec.~\ref{sec:results}, using
bootstrapped training data in Sec.~\ref{sec:results_boot}, a
truncated loss without fixed stopping condition in Sec.~\ref{sec:results_trunc}, and a loss that overrides the stop condition in Sec.~\ref{sec:results_mod}. In
Appendix~\ref{sec:discformer} we provide additional information on how
to improve the accuracy of the generative transformer through
including a classifier in the training, in the spirit of a GAN.

\section{Autoregressive jet radiation}
\label{sec:basics}

Given that jet radiation in QCD is described by universal splitting
kernels and well-defined scalings in the number of jets, we will train
an autoregressive transformer with a factorized likelihood loss to
generate QCD jet radiation.  The ultimate goal is to show that the
transformer not only describes jet radiation to a number of jets
represented in the training data, but that it can extrapolate to
larger jet counts than seen during training.

We first remind ourselves of universal splittings in QCD and the
typical scaling in the number of produced jets. We will then motivate
our $Z$+jets dataset, exhibiting the universal so-called staircase
scaling. To train a generative network we first derive a factorized
phase space probability and then encode it in a loss function for an
autoregressive transformer.

\subsection{QCD jet radiation}
\label{sec:basics_qcd}

Collinear parton splittings in the initial or final states are the
backbone of QCD predictions for hadron colliders. Their universal
nature is the basis of parton densities, parton showers, and jet
radiation, and it defines the structure of LHC
events~\cite{Ellis:1996mzs,Plehn:2009nd,Campbell:2017hsr}. A
challenging consequence of collinear splittings is that any hard
scattering process is accompanied by a variable number of jets in the
final state, as described by jet radiation and parton showers in the
multi-purpose event
generators~\cite{Bellm:2015jjp,Sjostrand:2014zea,Alwall:2014hca,Sherpa:2024mfk}.
Combining parton shower and hard matrix element predictions is the
theory basis for the entire precision physics program at the
LHC~\cite{Catani:2001cc,Mangano:2002ea,Frixione:2007vw,Bothmann:2023ozs}.

\subsubsection*{Universal autoregressive structure}

The physics background of our paper is the universal nature of jet
radiation from collinear splittings, reflecting the collinear
factorization of the matrix element and the phase space. It allows us
to generate events with $n+1$ final-state jets from events with $n$
final-state jets. For final state radiation this factorization is
schematically written as
\begin{align}
\sigma_{n+1} \sim \int \frac{dp^2}{p^2} dz \ \frac{\alpha_s}{2\pi} P(z) \sigma_n \; ,
\label{eq:split}
\end{align}
where $p^2$ is the invariant mass of the splitting parton, $z$ is the
momentum fraction carried out of the hard process $\sigma_n$, and
$P(z)$ are the universal collinear splitting kernels. In the initial
state, this factorization is the basis of the DGLAP equation with the
subtracted versions of the same collinear splitting kernels.

The iterative structure of Eq.\eqref{eq:split} allows us to simulate
parton splittings as Markov processes, and it also allows us to
describe the underlying densities in an approximately factorized
form. Such a factorized density is most efficiently generated by an
autoregressive structure. The key ingredients are the perturbative QCD
splitting functions and the non-splitting probability, referred to as
Sudakov factor.

The actual simulation of, approximately, collinear jet radiation is
not expected to be exact: first, we need to generate final transverse
momenta for the radiated partons while keeping transverse momentum
conservation~\cite{Hoche:2024dee}; second, we need to correct for
color and spin correlations~\cite{Buckley:2014fqa}; finally, the
structure of successive $(1 \to 2)$-splittings might not be
sufficient for the LHC
precision~\cite{vanBeekveld:2023ivn,vanBeekveld:2024wws}.
Nevertheless, the form of Eq.\eqref{eq:split} suggests that in QCD
events with increasing number of jets can be derived from a simple
iterative pattern, and such a pattern can in principle be learned and
extrapolated by a neural network with the right (autoregressive)
architecture.

\subsubsection*{Jet rate scaling}

The number of radiated jets in LHC events does not
follow a universal distribution. However, we can derive two distinct
patterns. Both are defined in terms of the ratio of $(n+1)$-jet to
$n$-jet events or in terms of the fraction of events with $n$ jets,
\begin{align}
 R_{(n+1)/n} 
 = \frac{\sigma_{n+1}}{\sigma_n}
 \qquad \text{and}  \qquad 
 P(n) = \frac{\sigma_n}{\sigma_\text{tot}}
 \qquad \text{with} \qquad 
 \sigma_\text{tot} = \sum_{n=0}^\infty \sigma_n \; .
\label{eq:njet}
\end{align}    
The ratios and the probabilities depend on kinematic cuts regularizing
the soft and collinear divergences, typically the minimum transverse
momentum of the counted jets, $p_{T,\text{min}}$.
\begin{enumerate}
\item 
The first pattern, Poisson scaling, implies in terms of the
expectation value $\bar{n}$,
\begin{align}
R_{(n+1)/n} 
= \frac{\bar{n}}{n+1} 
\qquad \Leftrightarrow \qquad 
P(n) = \frac{\bar{n}^n e^{-\bar{n}}}{n!} \; .
\label{eq:def_scaling}
\end{align}
At colliders, it occurs for processes with large splitting
probabilities and large scale differences, for instance multi-jet
production in $e^+ e^-$ collisions.

\item
We focus on the alternative staircase
scaling~\cite{Ellis:1985vn,Berends:1989cf,Berends:1990ax} with
\begin{align}
R_{(n+1)/n} 
= e^{-b} 
\qquad \Leftrightarrow \qquad 
P(n \geq n_\text{min}) = e^{- b n_\text{min}} \; 
\; .
\label{eq:staircase}
\end{align}
While the ratio $e^{-b}$ is the same for the exclusive and inclusive
jet counts, the probability only has a simple form for the inclusive
jet count, classifying events with $n_\text{min}$ jets or more. We can
use the universal scaling to relate $P(n)$ to a successive or
conditional probability
\begin{align}
  P(n+1|n) = R_{(n+1)/n} \; .
\label{eq:cond_scaling}
\end{align}
\end{enumerate}
At colliders, staircase scaling is predicted for smaller splitting
probabilities and democratic scales~\cite{Gerwick:2011tm}. In that
case, the jet count distributions can be derived from QCD using
generating functionals~\cite{Gerwick:2012hq}. For final state
radiation we quote the scale-dependent result
\begin{align}
 R_{(n+1)/n} 
 = 1 - \tilde{\Delta}_g(Q^2) \; ,
\end{align}
with a modified Sudakov factor or non-splitting probability
\begin{align}
 \tilde{\Delta}_g(Q^2) 
 &= \exp \left[ - C_A \int_{Q_0^2}^{Q^2} dt 
 \frac{\alpha_s(t)}{2\pi t} 
 \left( \log \frac{t}{Q_0^2}- \frac{11}{6} \right)
 \right] \; .
\end{align}    
To leading-log level the integrand is the QCD splitting function in
the collinear approximation. This QCD derivation of staircase scaling
requires democratic scales $Q^2/Q_0^2 \sim \mathcal{O}(1)$.

At the LHC the standard example is weak boson production with jets,
\begin{align}
pp \to Z + n \; \text{jets}
\qquad \text{with} \qquad n = \{ 0,1,2,3,... \}
\end{align}
Because the two scaling patterns are different, we will limit
ourselves to learning and generating staircase scaling from datasets
described by universal collinear radiation.

\subsection{\texorpdfstring{$Z$}{Z} + jets dataset}
\label{sec:basics_data}

\begin{figure}[b!]
    \centering
    \includegraphics[width=0.42\linewidth]{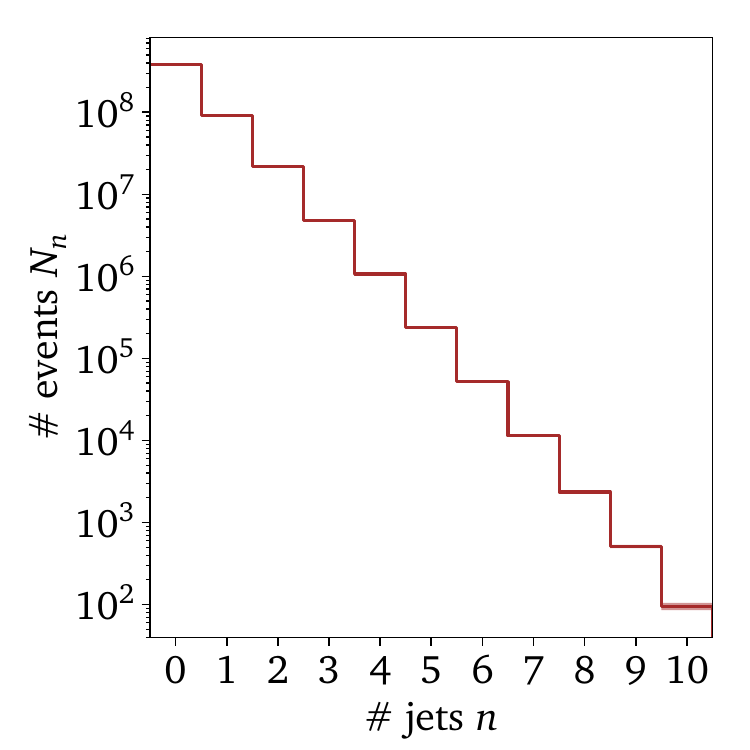}
    \hspace{0.6cm}
    \includegraphics[width=0.42\linewidth]{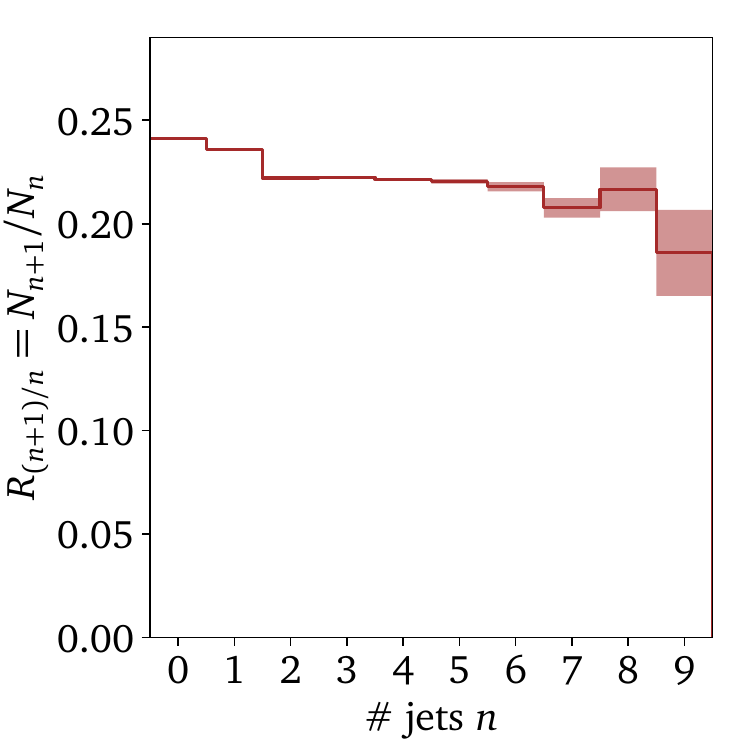}
    \caption{Staircase scaling of the number of jets in our $pp\to
      Z+n\;\text{jets}$ dataset. We show statistical uncertainties and
      use Gaussian error propagation to estimate the uncertainties for the
      ratio $R_{n+1/n}$.}
    \label{fig:jet_scaling}
\end{figure}

We follow the above motivation and
Refs.~\cite{Butter:2021csz,Butter:2023fov} by generating $Z$ bosons decaying to muons in association with a variable number of
jets. Unlike for earlier studies, we include higher jet multiplicities
to provide a challenge for the transformer
\begin{align}
    p p \to Z_{\mu\mu} + \{ 0,\,\cdots,10 \} \; \text{jets}.
\end{align}
We use \textsc{MadGraph5\_aMC\_@NLO}v3.5.1~\cite{Alwall:2011uj} to generate 500M $pp$ events at a center-of-mass energy of $\sqrt{s}=13\;\mathrm{TeV}$,
including ISR and parton shower with \textsc{Pythia}8~\cite{Sjostrand:2014zea}, using CKKW
merging~\cite{Catani:2001cc} and hadronization, but no pile-up. The jets are defined with
\textsc{FastJet}v3.3.4~\cite{Cacciari:2011ma} using the anti-$k_T$ algorithm~\cite{Cacciari:2008gp} with $R=0.4$ and the basic cuts
\begin{align}
p_{T,j} > 20~\gev 
\qquad \text{and} \qquad 
\Delta R_{jj} > 0.4.
    \label{eq:jet_cuts}
\end{align}
The muons and jets are both ordered in descending transverse momentum. Our phase
space dimensionality is three per muon and four per jet. Momentum
conservation is not guaranteed, because some final-state particles
might escape for instance the jet algorithm. The distribution of the
number of jets and the corresponding ratios $R_{(n+1)/n}$ are shown in
the two panels of Fig.~\ref{fig:jet_scaling}. We observe an
approximately constant ratio for most of the spectrum, confirming a
staircase scaling as defined in Eq.\eqref{eq:staircase}. Towards large
numbers of jets we start encountering statistical limitations as well
as phase space limitations.

Of our 500M events we use 80\% for training, 10\% for validation, and
10\% for testing. The number of events per jet multiplicity is given
in Tab.~\ref{tab:dataset_sizes}. To avoid being entirely dominated by
low-multiplicity events, we cap the number of events with $n=0,1,2$ to match
the number of events with $n=3$.

For the jet momenta, we use a minimal
preprocessing~\cite{Butter:2021csz,Butter:2023fov}, where each
particle $i$ is represented in standard jet coordinates
\begin{align}
    \left\{ \; ( p_T, \eta, \phi, m)_i \; \right\} \; .
\label{eq:def_obs}
\end{align}
We enforce the $p_T$ cuts in Eq.\eqref{eq:jet_cuts} using the
transformation $\log (p_T - p_{T,\text{min}})$, which maps allowed
transverse momenta to the full real line and leads to an approximately
Gaussian shape. The jet mass is encoded as $\log m$. We express angles
$\phi$ relative to the leading muon and apply a special treatment
described in Section~\ref{sec:architecture} to reflect the
periodicity. We suppress the leading muon $\phi$ angle due to the global rotation symmetry. Finally, we standardize all phase space variables except
$\phi$ as $(x - \bar{x})/\sigma(x)$. For 10 jets this phase space is 45-dimensional.

\begin{table}[t]
    \centering
    \begin{small} \begin{tabular}{rccccccccccc}
        \toprule
        Number of jets   & 0     & 1   & 2   & 3    & 4    &    5 &   6 &   7 &    8 &   9 & 10 \\
        \midrule
        Number of events & 380M  & 91M & 21M & 4.7M & 1.1M & 230k & 52k & 11k & 2.3k & 510 & 95 \\
        Cap & 4.7M  & 4.7M & 4.7M & - &  - & - & - & - & - & - & - \\
        \bottomrule
    \end{tabular} \end{small}   
    \caption{Event counts for our simulated $Z$+jets dataset. When
      training networks, we cap the size of the 0,1,2-jet subsets.}
    \label{tab:dataset_sizes}
\end{table}

\subsection{Factorized probability} 
\label{sec:likelihood}

Following the discussion in Sec.~\ref{sec:basics_qcd}, QCD jet
radiation has two features that make it an attractive target for
autoregressive generative networks: the universal splitting kernels
and the jet ratio patterns. In case of staircase scaling the ratios of
exclusive and inclusive jet rates are also
universal. Equation~\eqref{eq:split} suggests that the phase space
density for an event $x$ can be constructed as a product of
conditional distributions, each taking the form
\begin{align}
    p(x_i|x_{1:i-1}) = 
  \pkin(x_i|x_{1:i-1}) \;
  \psplit(x_{1:i-1}) \; ,
  \label{eq:kin_rad}
\end{align}
where we denote by $x_{1:i-1}$ the sequence of progenitor partons $x_1,\,\dots,
x_{i-1}$. For particle $i$, \pkin encodes the kinematics, conditional
on the probability \psplit that it will be radiated. Both
probabilities are conditioned on the full previous sequence of
particles $x_{1:i-1}$. This dependence on the complete previous sequence is necessary to describe non-Markovian corrections to the universal nature of splitting kernels.

Approximate universality of the splitting kernels and jet ratios
translates to universality of \pkin and \psplit respectively. This
raises the possibility that, given the right architecture, we can
train a neural network to extrapolate QCD jet radiation patterns in
analogy to a collinear parton shower Monte Carlo approach.

Using the conditional probabilities in Eq.\eqref{eq:kin_rad} we can
build the likelihood of an $n$-jet event,
\begin{align}
    p(x_{1:n}) 
    &= \left[ \prod_{i=1}^n p(x_i|x_{1:i-1})\right] \;
    \big[ 1-\psplit(x_{1:n})\big] \notag \\
    &= \left[ \prod_{i=1}^n \pkin(x_i|x_{1:i-1}) \right] \;
       \left[ \prod_{i=1}^n \psplit(x_{1:i-1}) \right] \;
    \big[ 1-\psplit(x_{1:n})\big] \;,
    \label{eq:likelihood0}
\end{align}
where the last term gives the probability that there are no further
splittings and the event is complete. In QCD language it corresponds
to a Sudakov factor.  In accordance with Eq.\eqref{eq:def_obs}, the
phase space probability $p(x_{1:n})$ has a well-defined dimensionality
$4^n$. It is normalized both as a continuous distribution over $x_i$
and a categorical distribution over $n$,
\begin{align}
    \sum_{n=1}^\infty \int\mspace{-8mu}\text{d}x_1\dots\text{d}x_n\,p(x_{1:n}) = 1\;.  
\end{align}

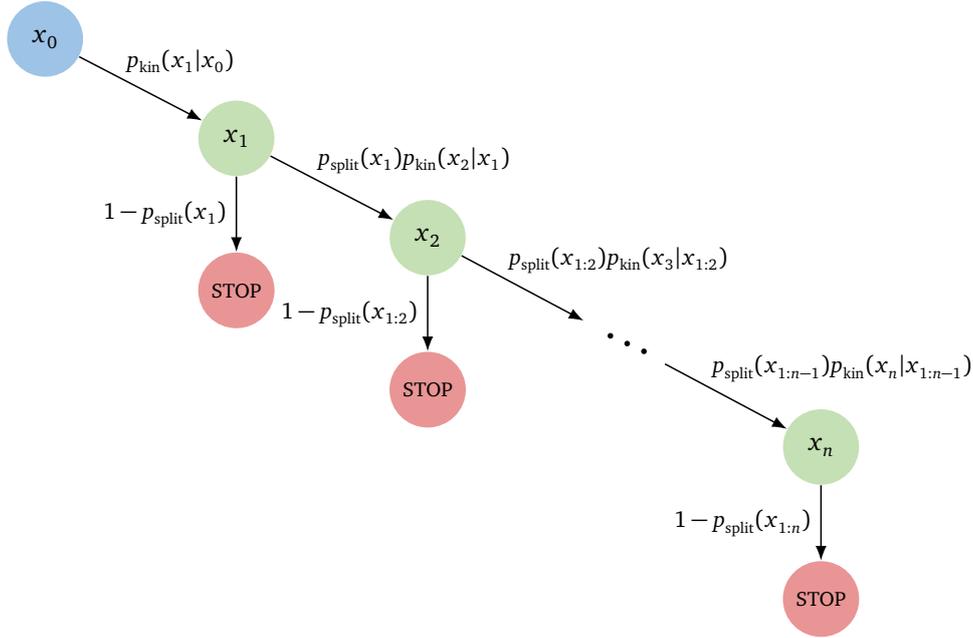
\begin{figure}[t]
    \centering
    \begin{tikzpicture}[font=\footnotesize]

\node (x0) [circle, fill=Bcolor, minimum width=1cm] {\small $x_0$};

\node (x1) [circle, fill=Gcolor, minimum width=1cm, below right = 0.6cm and 1.8cm of x0] {\small $x_1$};
\node (s1) [circle, fill=Rcolor, minimum width=1cm, below = 1cm of x1] {\scriptsize STOP};

\node (x2) [circle, fill=Gcolor, minimum width=1cm, below right = 0.6cm and 1.8cm of x1] {\small $x_2$};
\node (s2) [circle, fill=Rcolor, minimum width=1cm, below = 1cm of x2] {\scriptsize STOP};


\node (dotsb) [circle, fill=white, minimum width=1.2cm, below right = 0.6cm and 1.8cm of x2] {};
\node (dots) [rotate=155, anchor=center] at (dotsb) {\LARGE $\dots$};
\node (xn) [circle, fill=Gcolor, minimum width=1cm, below right = 0.6cm and 1.8cm of dotsb] {\small $x_n$};
\node (sn) [circle, fill=Rcolor, minimum width=1cm, below = 1cm of xn] {\scriptsize STOP};

\draw [arrow] (x0) -- node [above right,pos=0.3,anchor=west, yshift=0.2cm] {$\pkin(x_1|x_0)$} (x1);
\draw [arrow] (x1) -- node [above right,pos=0.3,anchor=west, yshift=0.2cm] {$\psplit(x_1)\pkin(x_2|x_1)$} (x2);
\draw [arrow] (x1) -- node [left,midway] {$1-\psplit(x_1)$} (s1);
\draw [arrow] (x2) -- node [left,midway] {$1-\psplit(x_{1:2})$} (s2);
\draw [arrow] (x2) -- node [above right,pos=0.3,anchor=west, yshift=0.2cm] {$\psplit(x_{1:2})\pkin(x_3|x_{1:2})$} (dotsb);

\draw [arrow] (dotsb) -- node [above right,pos=0.3,anchor=west, yshift=0.2cm] {$\psplit(x_{1:n-1})\pkin(x_n|x_{1:n-1})$} (xn);
\draw [arrow] (xn) -- node [left,midway] {$1-\psplit(x_{1:n})$} (sn);
\end{tikzpicture}
    \caption{Probability tree for variable-length event generation. To
      disallow empty events, we assign $\psplit(x_0) = 1$.}
    \label{fig:stoptree}
\end{figure}

As illustrated in Fig.~\ref{fig:stoptree}, the generative process can
be visualized as a binary probability tree with a Sudakov stop if no
further splitting happens. The combination of \psplit for a splitting
or $(1-\psplit)$ for no splitting is described by a Bernoulli
distribution $\pbin$ with expected splitting probability \psplit
\begin{align}
    \pbin\left(y| \psplit\right) = \psplit^y(1-\psplit)^{1-y}
    \quad \text{with} \quad 
    y\in\{0,1\}, \; \psplit\in [0, 1]\;.
    \label{eq:bernoulli}
\end{align}
It allows us to unify the factors \psplit and $1-\psplit$ in a
completely factorized likelihood
\begin{align}
    p(x_{1:n}) = 
    \prod_{i=1}^n \pkin(x_i|x_{1:i-1}) \; 
    \prod_{i=0}^n \pbin\big(1-\delta_{in}|\psplit(x_{1:i})\big)\;.
    \label{eq:full_factorization}
\end{align}
The Kronecker delta $\delta_{in}$ assigns the splitting label zero for the
$n$\textsuperscript{th} particle and one otherwise. By keeping the
full conditioning on $x_{1:i}$, this likelihood is completely general
and can capture non-universal correlations. This is important when we
describe full events, including the hard process. For
$Z_{\mu\mu}+$jets events, we also include the muons in the sequence, but explicitly set their splitting probabilities to one instead of learning them. In addition, we use an additional one-hot encoded network input to distinguish the two types of muons and the jets.

Similarly to the decomposition of the event likelihood $p(x_{1:n})$, we autoregressively factorize the likelihood of
individual particles $\pkin (x_{i+1}|x_{1:i})$ in terms of their
components. The ordering of components can affect the network
performance~\cite{Butter:2023fov}, however for such small sequences
this effect is negligible. The elements of the sequence are
one-dimensional, and we parametrize their distributions with mixtures
\begin{align}
    p_\text{kin}(x_{i+1}|x_{1:i}) 
    = \ &p_\text{GM}(p_{T,i+1}| x_{1:i})\, p_\text{vMM}(\phi_{i+1}|x_{1:i},p_{T,i+1})\,\notag \\
    \times &p_\text{GM}(\eta_{i+1}|x_{1:i},p_{T,i+1},\phi_{i+1})\,  p_\text{GM}(m_{i+1}|x_{1:i},p_{T,i+1},\phi_{i+1},\eta_{i+1})\;.
    \label{eq:likelihood2}
\end{align}
We use Gaussian mixtures $p_\text{GM}$ for non-periodic variables and
von Mises mixtures $p_\text{vMM}$ for the periodic variable
$\phi$. Since it is straightforward to sample and to compute densities with these component-wise mixtures, the same is true for the full event likelihood. We do not generate the fixed muon mass in $Z_{\mu\mu}+$jets events. Periodic likelihoods for angular variables inform the network
about this geometric information and therefore improve the
performance. This has been previously shown for normalizing
flows~\cite{Ackerschott:2023nax} and conditional flow
matching~\cite{Spinner:2024hjm}.

In contrast to the autoregressive structure of $p(x_{1:n})$ in
Eq.\eqref{eq:likelihood0}, Eq.\eqref{eq:likelihood2} is not inspired
by physics and other choices are possible. Examples from the
literature are categorical distributions over bins (which suffer from
limited
resolution)~\cite{Finke:2023veq,Golling:2024abg,Butter:2023fov},
normalizing flows~\cite{Heimel:2023mvw, Leigh:2024ked}, and
conditional flow matching~\cite{Heimel:2023mvw, Leigh:2024ked}.

We emphasize that this factorized likelihood, built to describe an
autoregressive generation, generalizes the usual factorization
$p(x_{1:n})=p(x_{1:i}|i)p(i)$~\cite{Butter:2021csz,Heimel:2023mvw} and
previous autoregressive approaches~\cite{Butter:2023fov}. Similar
generative approaches have been developed for jet constituent
generation~\cite{Birk:2024knn}, and a similar factorization for
density estimation has been studied in Refs.~\cite{Andreassen:2018apy,
  Andreassen:2019txo}.

\subsection{Autoregressive transformer}
\label{sec:architecture}

Starting from the physics-motivated factorization in
Eq.\eqref{eq:full_factorization}, we need to encode these densities
with variable-length inputs~$x_{1:i}$ using neural networks. A
transformer $f_\theta$ with causal attention mask will turn these
sequences into fixed-sized representations. We use a pre-layernorm
transformer decoder with GeLU activations, for more information see
App.~\ref{app:hyperparameters}, and decompose the transformer output
as
\begin{align} 
f_\theta{(x_{1:i})} 
&= (\rho_i, \nu_i) \in \mathbb{R}\times\mathbb{R}^d
\;.
\end{align}
The embedding dimension $d$ is a hyperparameter. The $\rho_i$
represent the splitting probabilities that parametrize the Bernoulli
distributions,
\begin{align}
    \rho_i \approx \psplit(x_{1:i})\;.
\end{align}
The embeddings $\nu_i$ similarly parametrize the kinematic
conditionals
\begin{align}
\pkin(x_i|\nu_{i-1})\approx \pkin(x_i|x_{1:i-1})\;.
\end{align}
For clarity, we always suppress the dependence of $\rho_i$ and $\nu_i$
on $x_{1:i}$ and on the transformer parameters $\theta$.

\subsubsection*{Loss and training}

The loss function of the autoregressive network is given by the
likelihood in Eq.\eqref{eq:full_factorization},
\begin{align}
    \loss_\text{like}
    &= - \XLangle \log p(x_{1:n}) \XRangle_{x\sim \pd} \notag \\
    &= - \XXLangle \sum_{i=1}^n \log \pkin(x_i|\nu_{i-1}) +\sum_{i=0}^n       \log \pbin(1-\delta_{in};  \rho_i) \XXRangle_{x\sim\pd}\;.    
    \label{eq:loss}
\end{align}
The first term is the usual likelihood loss for the kinematic
generative network. The second term is the standard binary cross
entropy. In our generative network it implicitly enforces the correct
event multiplicity through a splitting discriminator.

In Sec.~\ref{sec:results} we will consider modified training
strategies to extrapolate beyond the maximal multiplicity
$n_\text{max}$ of events contained in the training dataset. One
strategy is to modify the cross entropy part of the likelihood loss in
Eq.\eqref{eq:loss}, for example by removing the contribution from the
term with highest multiplicity $n_\text{max}$
\begin{align}
    \loss_\text{trunc} 
    = \XXLangle -\sum_{i=1}^n \log \pkin(x_i|\nu_{i-1}) -\sum_{i=0}^n (1-\delta_{in_\text{max}})\log \pbin(1-\delta_{in};  \rho_i) \XXRangle_{x\sim\pd}\; .
    \label{eq:trunc_loss2}
\end{align}
Using this loss, the splitting prediction for maximum-multiplicity
events, $\rho_{n_\text{max}}$, is not explicitly trained. Rather,
the weight sharing in the transformer allows correlations learned at
lower multiplicity to be recycled.

When training our transformers on the $Z+$jets dataset from
Sec.~\ref{sec:basics_data}, we use the Adam optimizer with constant
learning rate $3\times 10^{-4}$ and batch size 1024. The batches
contain events with different multiplicities following the distribution in the training data. The validation loss is
tracked every 5k iterations, and we restore the network from the
checkpoint with lowest validation loss after 200k iterations.

\subsubsection*{Sampling}

To generate full events $x$, we sequentially sample from the
likelihood described in Sec.~\ref{sec:likelihood}, as visualized in
Fig.~\ref{fig:stoptree}. We sample 10M events in total and split them
according to their multiplicities. This procedure generates samples
from the exact likelihood learned by the network, but does not give us
explicit control over the generated jet multiplicities. We decide on a
maximum number of jets, and discard events for which the transformer
predicts further splittings.

\subsubsection*{Bayesian network}

Because we hope to use the autoregressive transformer for
extrapolation beyond the jets present in the training data, we need to
quantify the uncertainty in the predicted phase space density. We
resort to Bayesian neural networks
(BNN)~\cite{bnn_early,bnn_early2,bnn_early3,deep_errors} as a way to
learn systematic and statistical uncertainties together with the mean
network predictions. These are a standard method in LHC physics, for
instance for amplitude regression~\cite{Badger:2022hwf},
calibration~\cite{Kasieczka:2020vlh,ATLAS:2024rpl}, and
classification~\cite{Bollweg:2019skg}. They can be generalized to the
density estimation aspect of generative
networks~\cite{Bellagente:2021yyh,Butter:2021csz,Butter:2023fov,Bieringer:2024nbc},
where they return an uncertainty on the unit event weight.

BNNs replace the network parameters $\theta$ by learnable
distributions $q(\theta )$, usually assumed to be uncorrelated
Gaussians. Their loss consists of a sampled likelihood term and a
regularization with a prior-width hyperparameter,
\begin{align}
    \loss_\text{BNN}
    = - \XLangle \log p(x) \XRangle_{x \sim \pd, \theta \sim q} + D_\text{KL}\left[q(\theta), p(\theta )\right]\;.
\end{align}
To evaluate the BNN we sample from the learned weight distributions,
in our case generating 10 samples, with a given number of 10M events each.

\section{Results}
\label{sec:results}

Even though we are interested in extrapolating towards unseen jet
numbers, we first benchmark the accuracy of our transformer in
Sec.~\ref{sec:results_no}. We also show how without modifications the
generative network does not actually extrapolate. For a successful
extrapolation we first use a bootstrap approach in
Sec.~\ref{sec:results_boot} and then show in Sec.~\ref{sec:results_trunc} and Sec.~\ref{sec:results_mod} how truncating or overriding the likelihood loss allows the network to generate larger jet
numbers than seen during training.

\subsection{Generating without extrapolation}
\label{sec:results_no}

\begin{figure}[t!]
    \centering
    \includegraphics[width=0.49\textwidth]{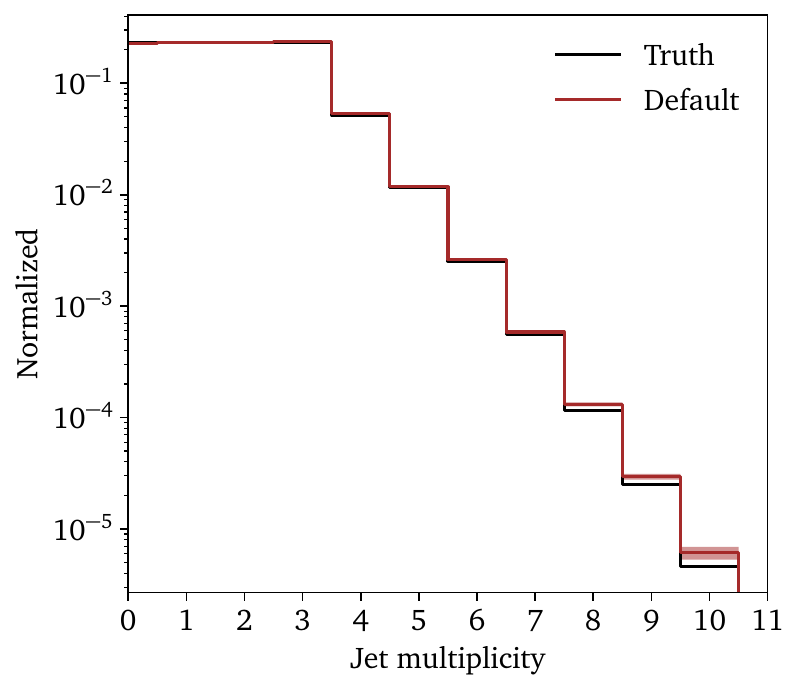}
    \hfill
    \includegraphics[width=0.49\textwidth]{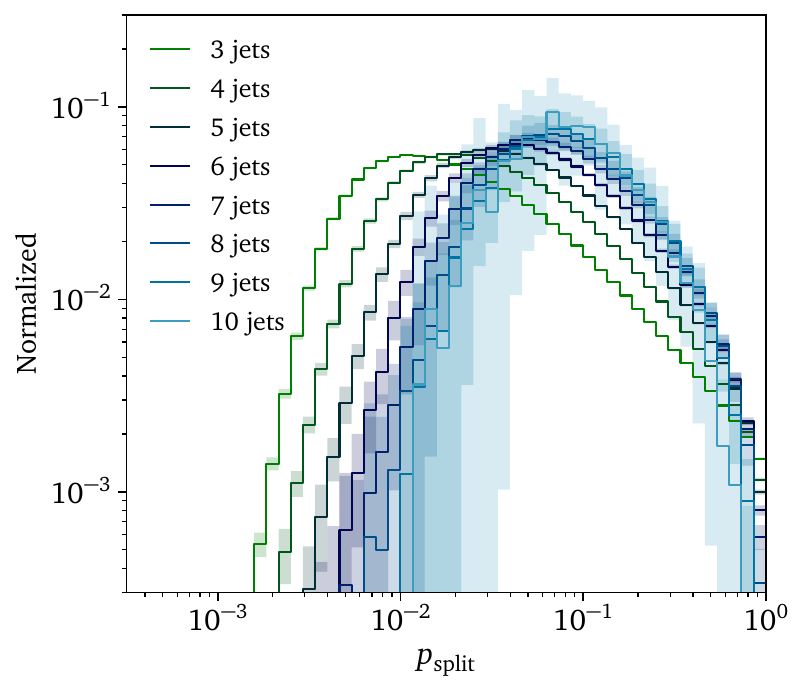} 
    \caption{Jet multiplicity distribution (left) and splitting probabilities (right) for samples generated with the transformer trained on the full dataset up to 10 jet events.}
    \label{fig:baseline_stop_convergence}
\end{figure}

\begin{figure}[b!]
    \includegraphics[width=0.325\textwidth, page=29]{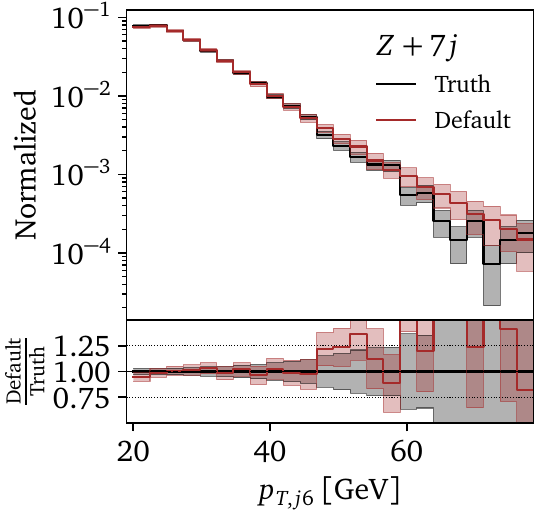} 
    \includegraphics[width=0.325\textwidth, page=33]{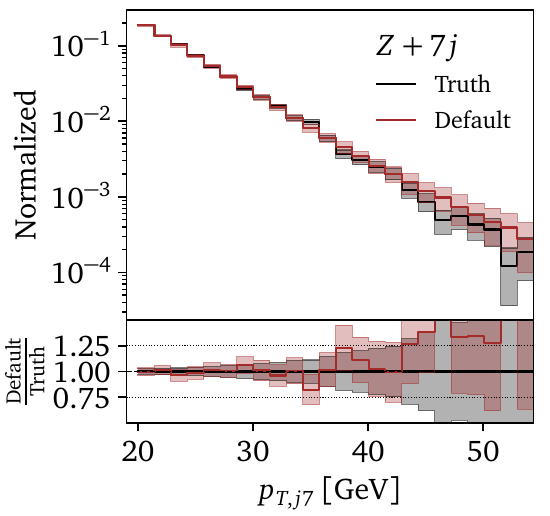} 
    \includegraphics[width=0.325\textwidth, page=4]{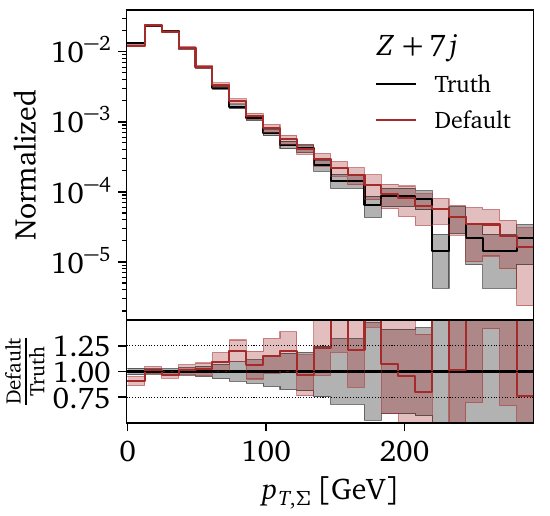} \\ 
    \includegraphics[width=0.325\textwidth, page=69]{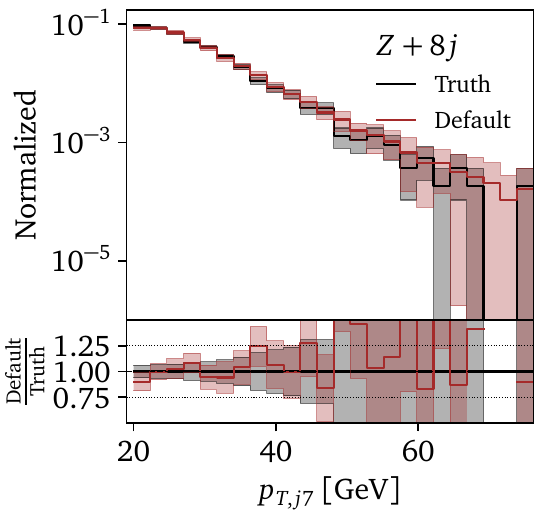}  
    \includegraphics[width=0.325\textwidth, page=73]{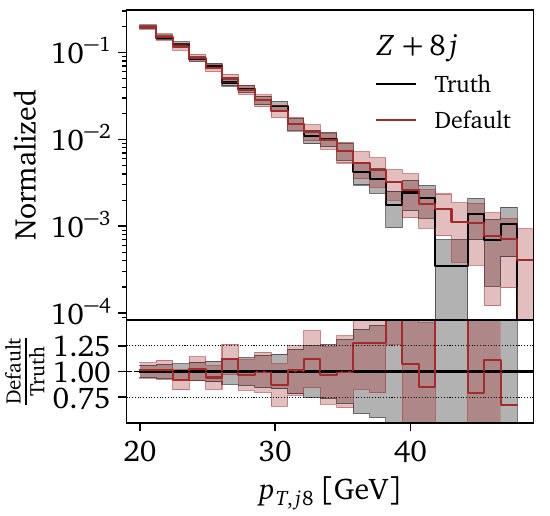}  
    \includegraphics[width=0.325\textwidth, page=8]{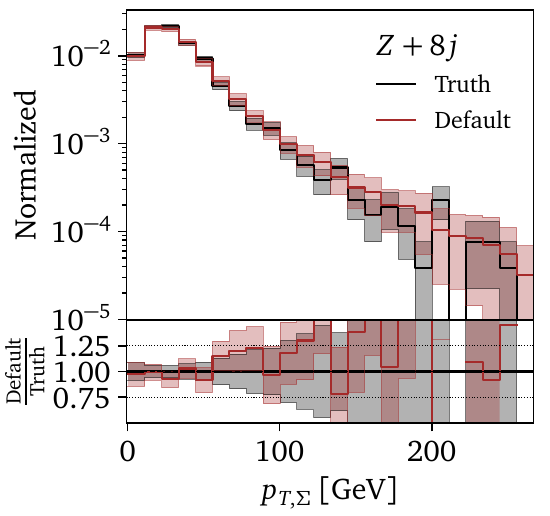} 
    \caption{Selection of features in $Z$ + 7 and 8-jet events for the
      generative network trained on the full dataset, including 7 and 8-jet events.}
    \label{fig:samples_baseline_new}
\end{figure}

We begin by demonstrating that our transformer learns the phase space
density precisely across event multiplicity. We train a Bayesian
version of the transformer using all $Z+n$ jet events, from the hard
process only, or $n=0$, up to $n=10$. We sample 10M events each from 10 BNN
predictions. The jet multiplicity distribution is shown in the left panel of Fig.~\ref{fig:baseline_stop_convergence}, showcasing that the generator can learn the universal staircase scaling. In Fig.~\ref{fig:samples_baseline_new}, we see that the
network reproduces the kinematic distributions with precision down to
the statistical uncertainty of the test set. The transverse component of the vector sum of all particle momenta, $p_{T,\Sigma}$, provides a sensitive test
of learned global correlation among all particles.
All deviations from the training data are captured by the Bayesian
uncertainty.

Next, we inspect whether the network has learned universal structure
in the probability to generate additional jets. In the right panel of Fig.~\ref{fig:baseline_stop_convergence} we show the distributions of
\psplit predicted during the autoregressive sampling steps.  We train
the network on the entire dataset, with up to 10~jets. We ignore the
learned $\psplit$ for the first two jets, because we manually
capped the number of training events for up to two jets, as shown in
Tab.~\ref{tab:dataset_sizes}. For more than 6 jets, the distribution stabilizes within
the Bayesian uncertainty band, indicating that between 6 and 10 jets
we do not observe a significant effect from the parton
densities~\cite{Gerwick:2012hq}.

\begin{figure}
    \centering
    \includegraphics[width=0.42\textwidth]{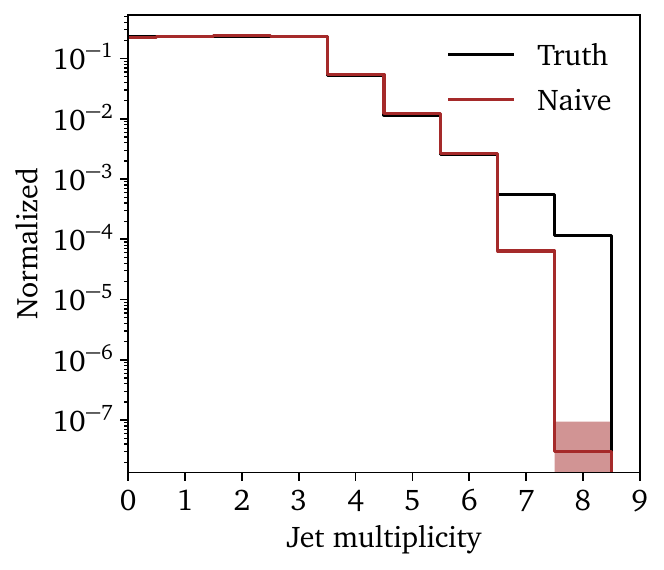}
    \includegraphics[width=0.42\textwidth]{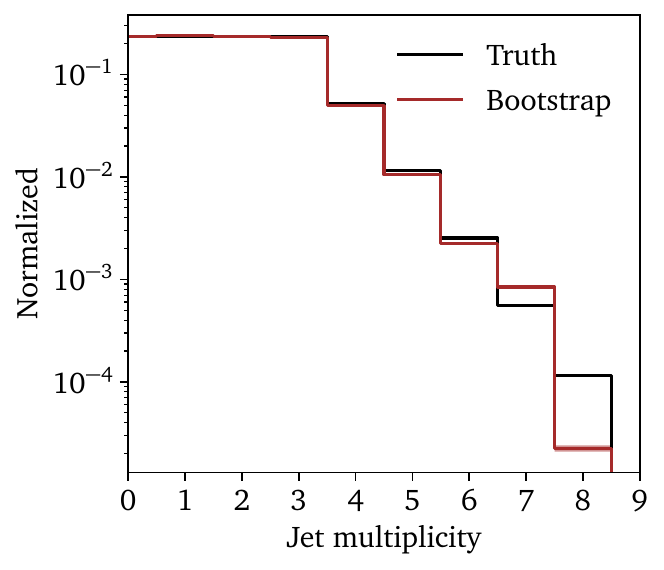}
    \hspace{0.6cm}
    \caption{Jet multiplicity distributions for samples generated with the transformer using the naive training (left) and bootstrapping (right).}
    \label{fig:baseline_asis_probstop}
\end{figure}

\subsubsection*{Naive extrapolation}

\begin{figure}[b!]
    \includegraphics[width=0.325\textwidth, page=29]{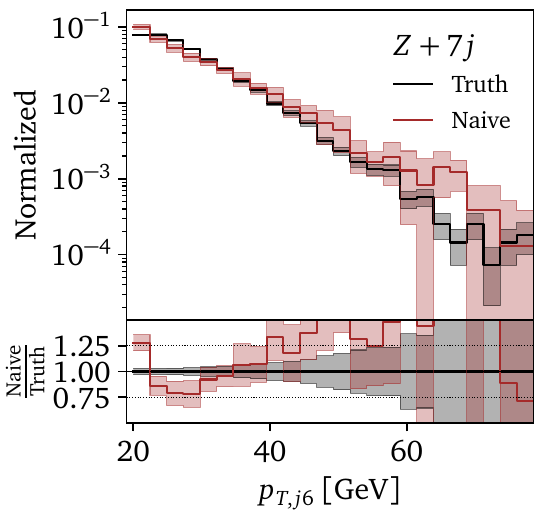} 
    \includegraphics[width=0.325\textwidth, page=33]{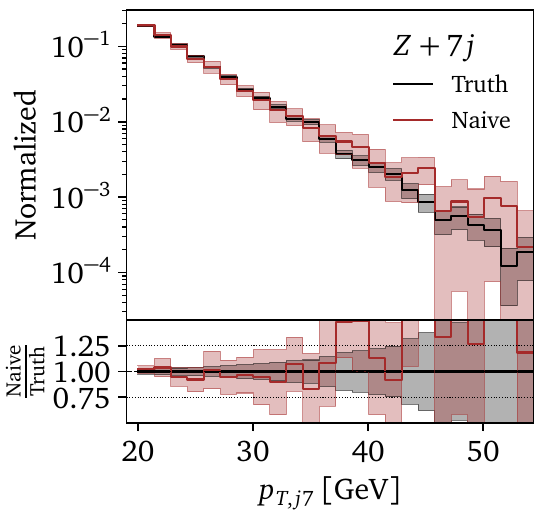} 
    \includegraphics[width=0.325\textwidth, page=4]{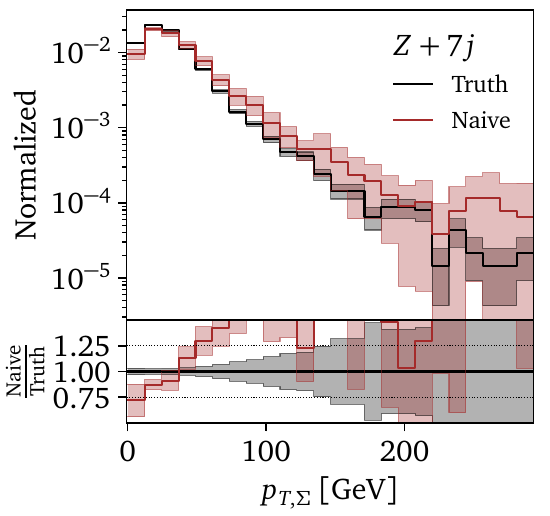} \\ 
    \includegraphics[width=0.325\textwidth, page=69]{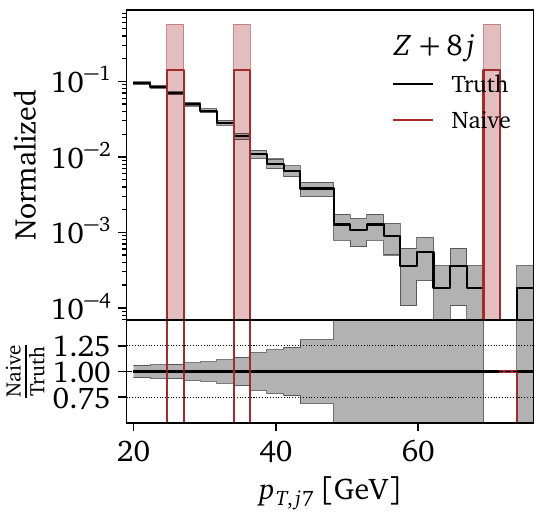}  
    \includegraphics[width=0.325\textwidth, page=73]{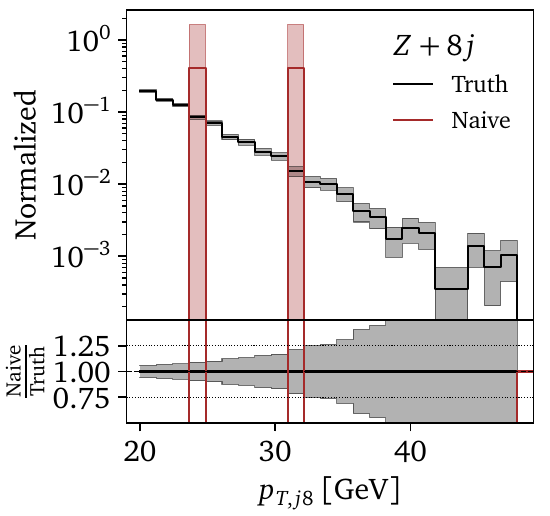}  
    \includegraphics[width=0.325\textwidth, page=8]{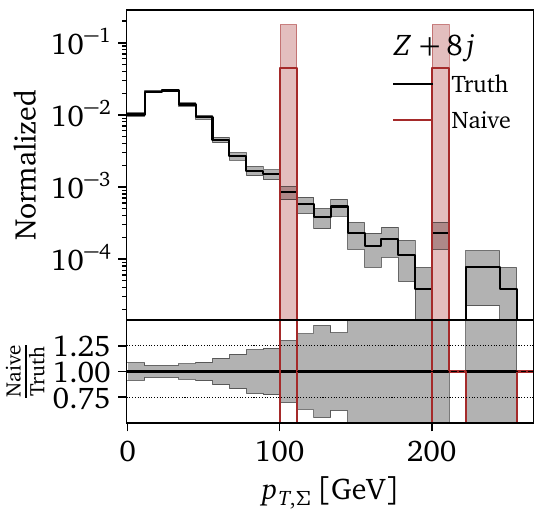} 
    \caption{Selection of features in $Z + 7$ and 8-jet events for a generator trained on up to 6-jet events.}
    \label{fig:samples_asis}
\end{figure}

Because the termination of the number of jets is implemented
probabilistically, we can naively extrapolate to higher jet numbers. For
instance, we can train the networks with up to 6 jets and assess the
small number of 7-jet and 8-jet events they generate. While the
quality of 7-jet and 8-jet events should be worse than for jet numbers
seen during training, we want to know if the transformer can leverage
universal properties of the QCD jet radiation. We show the generated
jet multiplicity distribution in the left panel of
Fig.~\ref{fig:baseline_asis_probstop}. Indeed, the network generates
events with more than 6 jets, albeit with much lower probability than
expected from staircase scaling.

For perfect training, we expect the rate for events with more jets
than the training set to approach zero. This is because the
transformer output $\rho_i$ is trained to match the probability that
another jet follows particle $i$, $\rho_i\approx \psplit(x_{1:i})$. In
a given training set, with maximum event length $n_\text{max}$, one
always has
\begin{align}
  \psplit(x_{1:n_\text{max}})=0 \; .
  \label{eq:hard_cond}
\end{align}
The optimal network would learn $\rho_{n_\text{max}}=0$, and the
transformer can ignore physical correlations. The reason we do not
observe exact zero splitting probabilities is that the weight sharing
in the autoregressive transformer imparts a bias to reuse the pattern
learned at low multiplicities.

Given the small but finite rate of 7-jet events generated through
naive extrapolation, we want to see if the transformer has generalized
the jet kinematics. In Fig.~\ref{fig:samples_asis}, we show the
kinematic features as for the extrapolated 7-jet events. Among 100M
generated events, the transformer only generates three 8-jet events,
so we cannot assess their quality. However, the 7-jet events
look qualitatively reasonable.  In particular, the slightly broken
transverse momentum conservation is reproduced with an accuracy
similar to the baseline in Fig.~\ref{fig:samples_baseline_new}.  Given
that the $p_T$ of the 7\textsuperscript{th} jet is approximately the
same scale as the level of momentum non-conservation, this is a
non-trivial result. It suggests that the transformer indeed
generalizes kinematics, and we should mainly address the learned jet
multiplicity.

\subsection{Extrapolation with bootstrap}
\label{sec:results_boot}

\begin{figure}[t]
    \centering
    \begin{tikzpicture}[font=\footnotesize]
\usetikzlibrary{decorations.pathreplacing,positioning,calc}

\node (start) [rectangle, align=center, rounded corners, fill=Bcolor, minimum width=1.7cm, minimum height=0.7cm] {start};
\node (warmup) [rectangle, align=center, rounded corners, fill=Gcolor, minimum width=1.7cm, minimum height=0.7cm, right = 1cm of start] {warm-up};
\draw [arrow] (start) -- node[yshift=0.3cm] {200k} (warmup);

\node (n1) [rectangle, align=center, rounded corners, fill=Gcolor, minimum width=0.3cm, minimum height=0.3cm, right=0.7cm of warmup] {};
\draw [arrow] (warmup) -- node[yshift=0.3cm] {50} (n1);

\node (newsample) [circle, fill=Gcolor, minimum size=3pt, right=1.3cm of n1] {};
\node (ndots0)  [left=0.005cm of newsample] {$\cdots$};
\draw [arrow] (n1) -- node[yshift=0.3cm] {50} (ndots0);

\node [draw, fill=Gcolor!50, rounded corners=1pt, minimum width=0.45cm, minimum height=0.45cm, inner sep=0pt]
  at ([xshift=-0.3cm,yshift=-0.45cm]warmup.south) {7j};
\node[draw, fill=Gcolor!50, rounded corners=1pt, minimum width=0.45cm, minimum height=0.45cm, inner sep=0pt]
  at ([xshift=-0.3cm,yshift=-0.95cm]warmup.south) {7j};
\node[draw, fill=Gcolor!50, rounded corners=1pt, minimum width=0.45cm, minimum height=0.45cm, inner sep=0pt]
  at ([xshift=+0.2cm,yshift=-0.95cm]warmup.south) {7j};

\node (n2) [rectangle, align=center, rounded corners, fill=Gcolor, minimum width=0.3cm, minimum height=0.3cm, right=0.7cm of newsample] {};
\draw [arrow] (newsample) -- node[yshift=0.3cm] {50} (n2);

\node (buffer) [
    draw, very thick, rectangle, align=center, rounded corners,
    fill=none, minimum width=2.9cm, minimum height=0.75cm,
    ] at ([xshift=-1.1cm, yshift=-1cm]newsample) {};

\path (warmup) -- (n1) coordinate[midway] (midWN);
\draw[arrow] (midWN |- buffer.north) -- (midWN);
\path (n1) -- (ndots0) coordinate[midway] (midN1dots0);
\draw[arrow] (midN1dots0 |- buffer.north) -- (midN1dots0);

\draw[arrow] (n1.south) -- (n1.south |- buffer.north);

\draw[arrow] (warmup.south) |- ([yshift=0cm]buffer.west);

\node (bufgrid) [inner sep=0pt, minimum width=2.975cm, minimum height=0.5cm, anchor=center]
      at (buffer.center) {};
    
\node (yellowsample) [draw, fill=Gcolor!50, rounded corners=1pt,
       minimum width=0.45cm, minimum height=0.45cm, inner sep=0pt]
      at (newsample |- bufgrid.center) {7j};

\coordinate (cell01) at ($ (yellowsample.north) $);
\draw[arrow] (newsample) -- ([yshift=0cm]cell01.north);
\path let \p1 = (bufgrid.center) in
  node (sampletoreplace) [draw, dashed, fill=Gcolor!20, rounded corners=1pt, minimum width=0.45cm, minimum height=0.45cm, inner sep=0pt]
    at ([xshift={-1.1cm}, yshift={0cm}] \p1) {7j}
  node[draw, fill=Gcolor!50, rounded corners=1pt, minimum width=0.45cm, minimum height=0.45cm, inner sep=0pt]
    at ([xshift={-0.575cm}, yshift={+0cm}] \p1) {7j}
  node[draw, fill=Gcolor!50, rounded corners=1pt, minimum width=0.45cm, minimum height=0.45cm, inner sep=0pt]
    at ([xshift={-0.05cm}, yshift={-0.cm}] \p1) {7j};
\node (ndots2) at ([xshift=-0.3cm]yellowsample.west) {$\cdots$};

\node (redsample) [draw, fill=Gcolor!50, rounded corners=1pt,
      minimum width=0.45cm, minimum height=0.45cm, inner sep=0pt]
  at ($ (sampletoreplace.center) - (0.cm,0.3cm) $) {7j};

\draw[arrow]
  (n2.south) |- ($(buffer.south)+(0,-0.4cm)$) -| (redsample.south);

\node (finish) [rectangle, align=center, rounded corners, fill=Bcolor, minimum width=1.7cm, minimum height=0.7cm, right = 0.7cm of n2] {finish};
\draw [arrow] (n2) -- node[yshift=0.3cm] {50} (finish);
\draw[decorate, decoration={brace, amplitude=6pt}]
  ([yshift=0.45cm]warmup.east) -- ([yshift=0.45cm]finish.west)
  node[midway, yshift=0.4cm] {200k};
\path (newsample) -- (n2) coordinate[midway] (midNewN2);
\path (n2) -- (finish)  coordinate[midway] (midN2Fin);
\draw[arrow] ($(buffer.east)+(0,0.12cm)$) -| (midNewN2);
\draw[arrow] ($(buffer.east)+ (0,-0.12cm)$) -| (midN2Fin);

\end{tikzpicture}
    \caption{Training workflow for extrapolation with bootstrap. The upper horizontal arrows denote iterations within the training process, and vertical arrows denote moving bootstrapped events to update the buffer and include buffered events in training. Additional information is given in the text.}
    \label{fig:bootstrap}
\end{figure}

\begin{figure}[b!]
    \includegraphics[width=0.325\textwidth, page=29]{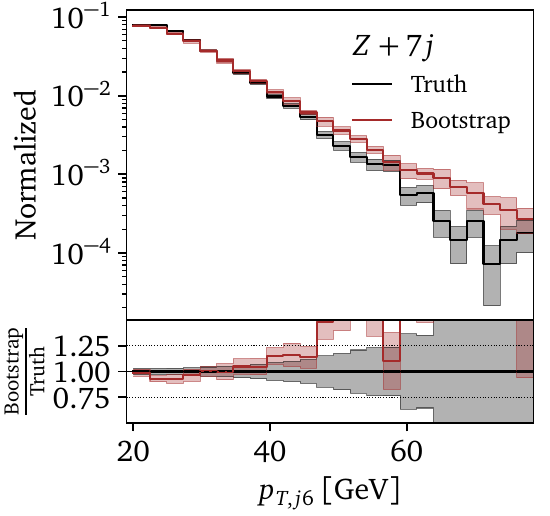} 
    \includegraphics[width=0.325\textwidth, page=33]{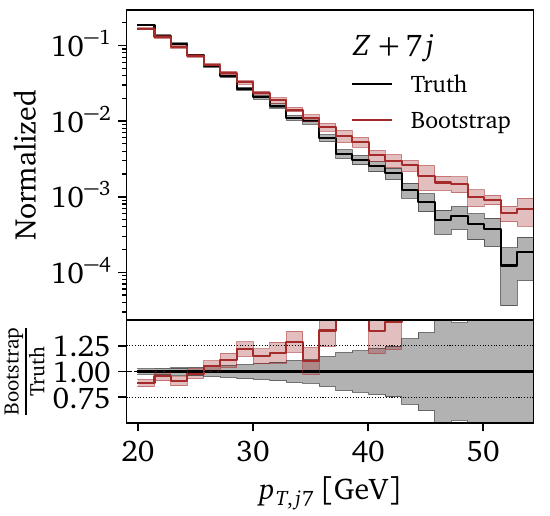} 
    \includegraphics[width=0.325\textwidth, page=4]{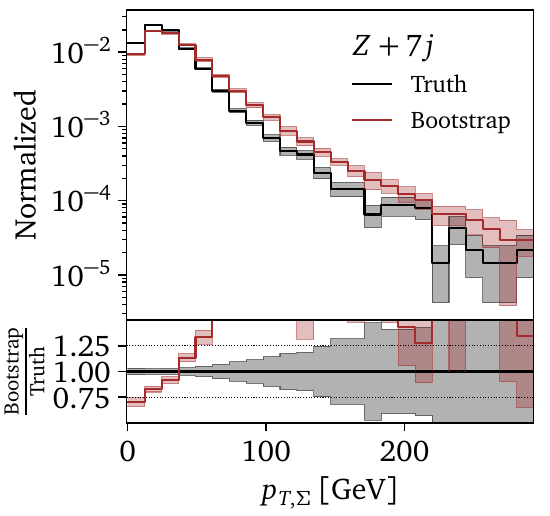} \\ 
    \includegraphics[width=0.325\textwidth, page=69]{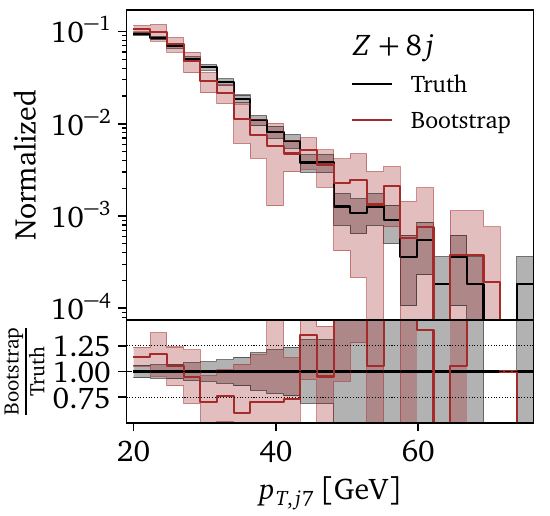}  
    \includegraphics[width=0.325\textwidth, page=73]{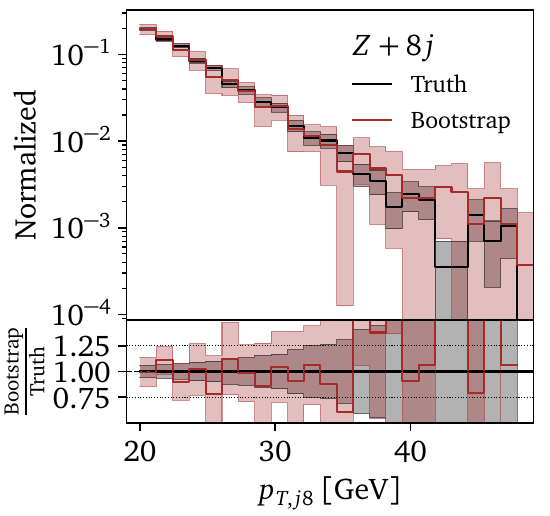}  
    \includegraphics[width=0.325\textwidth, page=8]{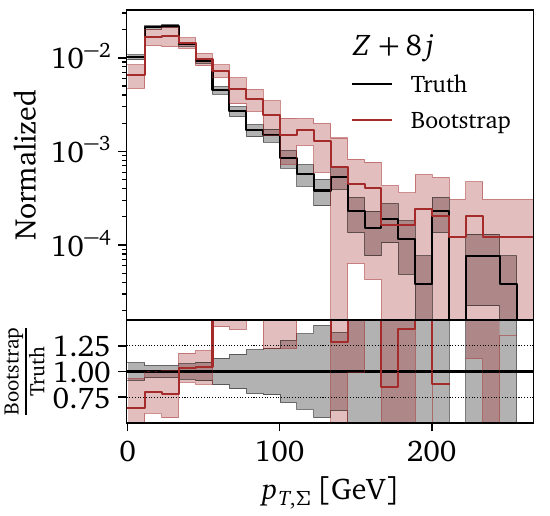} 
    \caption{Selection of features in $Z + 7$ and 8-jet events for a generator
      trained on up to 6-jet events using the bootstrap technique.}
    \label{fig:samples_bootstrap}
\end{figure}

A simple modification to increase the fraction of learned 7-jet events
is to bootstrap them, i.e. add generated 7-jet events to the training
data. This way, we dynamically break the condition
$\psplit(x_{1:n_\text{max}})=0$ of Eq.\eqref{eq:hard_cond} and allow
the network to adapt its multiplicity distribution. By repeating this
bootstrapping, we can also generate 8 jets and beyond. The fraction of
generated events introduced to the training dataset is a
hyperparameter. It controls the learned multiplicity distribution.

The training workflow is visualized in Fig.~\ref{fig:bootstrap}. We start to add bootstrapped events after a warm-up stage of 200k
iterations, corresponding to a full-length training in the approach of Sec.~\ref{sec:results_no}. Without
this warm-up stage, the network memorizes the poor-quality samples of
the freshly initialized network. After the warm-up, we generate a
buffer of 1k 7-jet events. For every generated batch we sample a new deterministic network from the learned weight distribution, making sure that we cover the full range of the weight posterior distribution. We then add a single
7-jet event to each batch of 1024 events, corresponding roughly to the
fraction of 7-jet events in the training dataset, and train for another 200k iterations with these settings. After every 50
iterations, we generate a batch of 32768 events, extract the 7-jet
events and add them to the buffer. Once the buffer contains 50k
events, we start to replace its oldest events with newly generated
events. This allows the network to dynamically adapt the quality of
7-jet events. We observe that the network has to be trained for a
sufficient amount of time in the bootstrapping mode to adapt to the
changed multiplicity distribution.

The obtained jet multiplicity distribution is shown in the right panel of Fig.~\ref{fig:baseline_asis_probstop}. We now get significantly more 7-jet
and 8-jet events, indicating that the network indeed adapts the
multiplicity distribution. The fraction of 8 jet events is significantly lower than in the training data, because we only bootstrap 7 jet events. The kinematical distributions of the generated events are shown in
Fig.~\ref{fig:samples_bootstrap}. They show that the bootstrapping generator yields valid kinematic configurations. However, there are deviations in the
kinematic features from the truth that are not covered by the Bayesian
uncertainty. We remark, however, that Bayesian uncertainties are not expected to cover out-of-distribution deviations.

\subsection{Extrapolation with truncated loss}
\label{sec:results_trunc}

\begin{figure}[b!]
    \centering
    \includegraphics[width=0.42\textwidth]{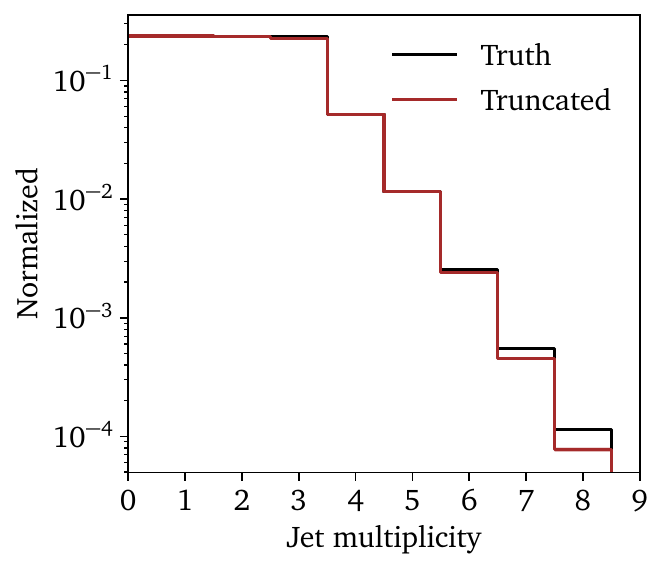}
    \hspace{0.6cm}
    \includegraphics[width=0.42\textwidth]{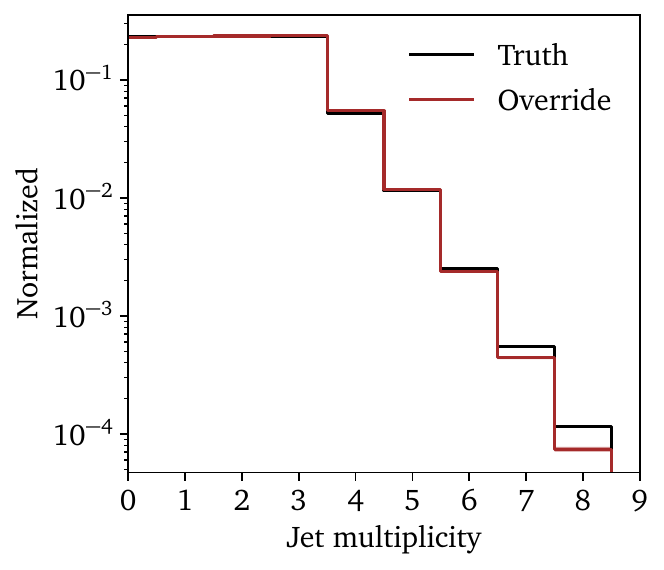}
    \caption{Jet multiplicity distributions learned using
      $\loss_\text{trunc}$ (left) and $\loss_\text{override}$ (right) trained on events with up to 6 jets.}
    \label{fig:probs_stop_loss}
\end{figure}

A complementary way to combat the suppression of events with more jets
than the training set is to modify the likelihood loss. As discussed
in Sec.~\ref{sec:results_no}, the cause of the suppression is the
constant $\psplit(x_{1:n_\text{max}})=0$ represented by a training
dataset with at most $n_\text{max}$ jets. A simple solution is to omit
the final Bernoulli contribution from the loss and truncate the loss
as described in Eq.\eqref{eq:trunc_loss2},
\begin{align}
    \loss_\text{trunc} = \left\langle -\sum_{i=1}^{n} \log \pkin(x_i|\nu_{i-1}) -\sum_{i=0}^n (1-\delta_{in_\text{max}})\log \pbin(1-\delta_{in};  \rho_i)\right\rangle_{x\sim\pd}\;,
    \label{eq:trunc_loss}
\end{align}
It differs from the complete likelihood loss of Eq.\eqref{eq:loss} in
the addition of the factor $1-\delta_{in_\text{max}}$ in front of the
Bernoulli component. Now, the splitting prediction for maximum-length
events, $\rho_{n_\text{max}}$, is not explicitly trained. Rather, the
weight sharing in the transformer allows correlations learned at lower
multiplicity to be recycled. When sampling a network trained in this
way, the splitting predictions beyond $n_\text{max}$ are pure
extrapolation.

\begin{figure}[t]
    \includegraphics[width=0.325\textwidth, page=29]{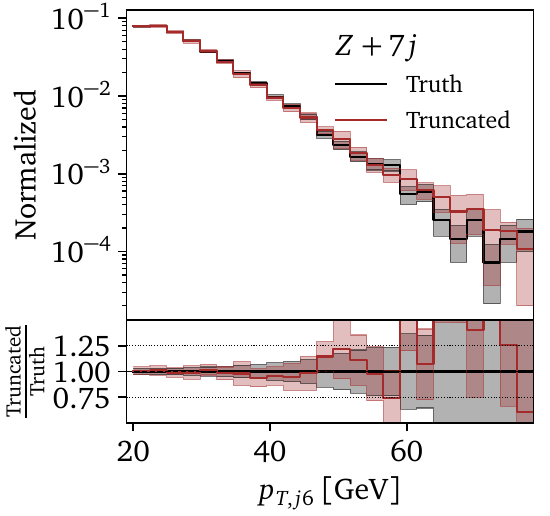} 
    \includegraphics[width=0.325\textwidth, page=33]{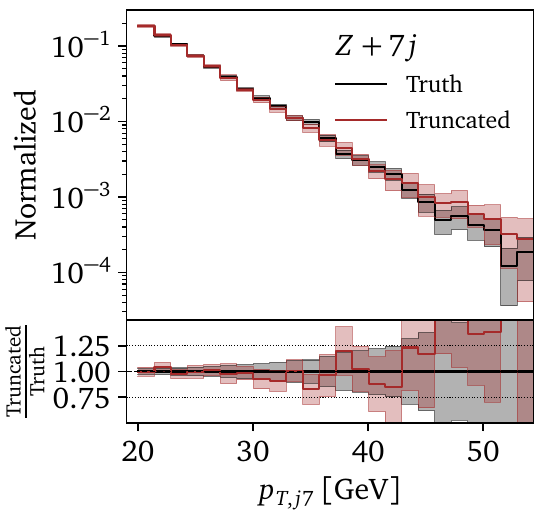} 
    \includegraphics[width=0.325\textwidth, page=4]{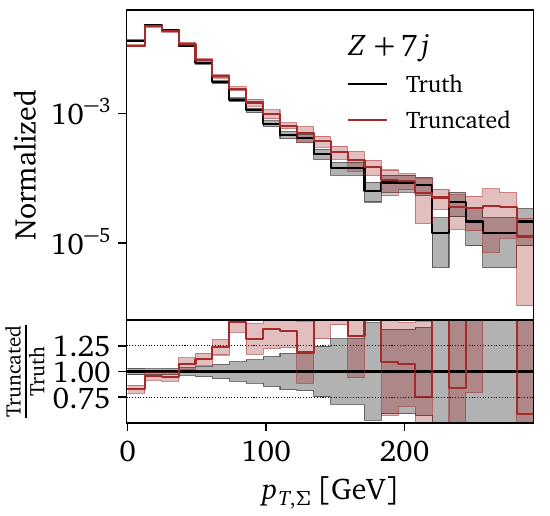} \\ 
    \includegraphics[width=0.325\textwidth, page=69]{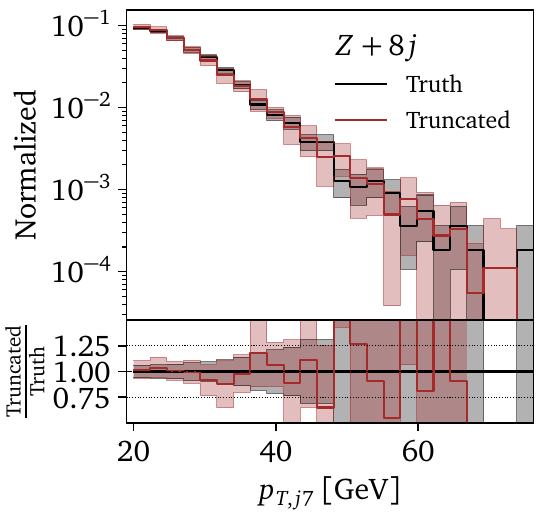}  
    \includegraphics[width=0.325\textwidth, page=73]{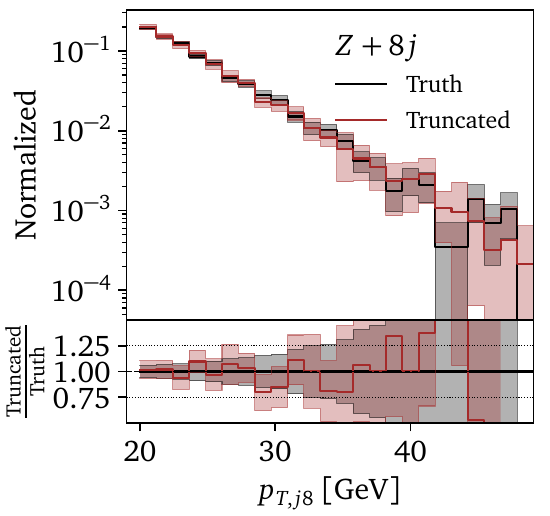}  
    \includegraphics[width=0.325\textwidth, page=8]{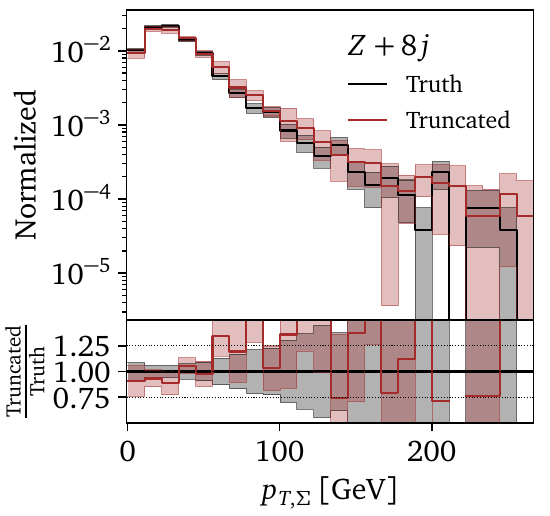} 
    \caption{Selection of features $Z + 7$ and 8-jet events, trained with the truncated loss.}
    \label{fig:samples_stop_trunc}
\end{figure}

Using the truncated loss, we again train a transformer on events with
up to 6 jets and again sample up to 8 jets. The generated
multiplicities are shown in Fig.~\ref{fig:probs_stop_loss} (left). Indeed,
the network learns and extrapolates the staircase scaling. We show the
extrapolated kinematic correlations in
Fig.~\ref{fig:samples_stop_trunc}. The only deviation exceeding the
BNN uncertainty is a slightly larger transverse momentum imbalance
than expected in 7-jet events. This result demonstrates that the
generative transformer described in Sec.~\ref{sec:basics} has learned
the universal pattern of jet radiation.

\subsection{Extrapolation with override}
\label{sec:results_mod}

In the previous section we have shown how truncating
the final Bernoulli term from the likelihood loss allows the network to generate high-quality
7-jet and 8-jet events. However, the extrapolation can be mis-calibrated. Because the transformer splitting
predictions $\rho_i$ are trained with a binary cross entropy, the 
optimal solution in terms of the non-splitting probability is the posterior
\begin{align}
    1-\rho_i 
    &\approx p(\text{stop at $i$}|x_{1:i})
    = \frac{p(x_{1:i}|\text{stop at $i$})}{\sum_{n\geq i} p(x_{1:i}|\text{stop at $n$})}
    \label{eq:stop_posterior}
\end{align}
assuming a uniform prior for simplicity. When training on a
dataset with maximum multiplicity $n_\text{max}$, the estimate is
biased since the sum over $n$ is missing terms above
$n_\text{max}$. The same effect causes the transformer to
stick to a constant splitting probability~$\rho_{n_\text{max}}=0$. 

In an alternative approach 
we show that transverse momentum conservation can be used as an extra 
handle on the posterior.
A violation of transverse momentum
conservation can be induced by removing particles beyond
the hard process and first $k$ jets. The spread in center of
momentum scales with $k$, so we can use transverse momentum conservation 
to statistically separate complete and incomplete
events. Secondly, we note that the $p_{x,\Sigma}$ and $p_{y,\Sigma}$
distributions are roughly Gaussians with zero mean, and hence fully
specified by their standard deviation.

\begin{figure}[t!]
    \centering
    \includegraphics[width=0.495\textwidth]{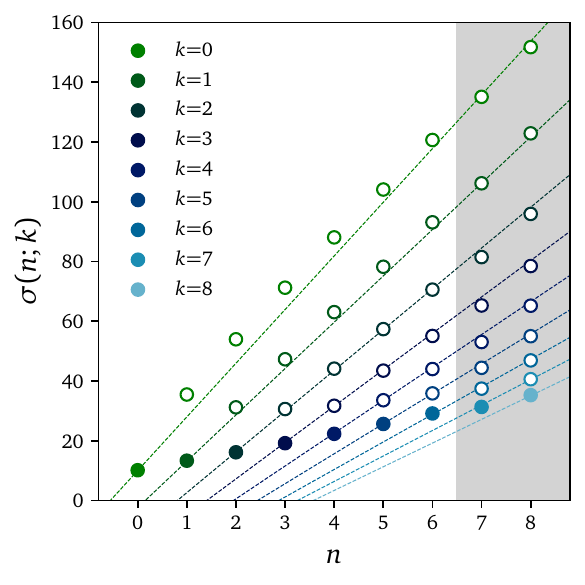}
    \caption{Standard deviations of  $p_{x,\Sigma}$ for the muons and first $k$ jets in $Z+n$-jet events. Filled circles indicate complete events, with $k=n$, while empty circles are incomplete. Points in the gray region are not used in any fit, but show the agreement of the extrapolation.}
    \label{fig:sigma_scaling}
\end{figure}

In Fig.~\ref{fig:sigma_scaling}, we show the widths of the
$p_{x,\Sigma}$ distributions as a function of the jet number, 
for complete events and for the hard process plus
$k$ jets. The widths obey an
approximately linear scaling when considering a fixed number of jets,
for complete events or otherwise. We can perform a linear fit to
estimate the standard deviations for higher-multiplicity events,
giving analytic expressions for the likelihoods in
Eq.\eqref{eq:stop_posterior}.
%
%
We arrive at
\begin{align}
    \sigma(n; k) &= (n - k)\,m_k 
 + \sigma(k; k)\;,  \notag \\[4pt]
 &\qquad\text{with}\quad
    \sigma(k; k) = 3.14 k + 9.97\;,  \notag \\
&\qquad\text{and}\quad     1/m_k = 0.0088 k + 0.056\;.    
\end{align}
The widths of completed events, $\sigma(k; k)$, are fit from the bottom row of filled circles in
Fig.~\ref{fig:sigma_scaling} up to $n=6$. The gradients $m_k$ of lines with constant $k$ are fit using events up to $n=5$.
The fits are shown as dotted lines in Fig.~\ref{fig:sigma_scaling}, and we see
that they extrapolate well to all partial $k$ values in 7-jet and 8-jet
events. Due to the rotation symmetry around the beam axis, the same
values hold for $p_{y,\Sigma}$ and we can assume that the joint
likelihoods are the product of 1D Gaussians.
\begin{align}
    p_\text{fit}(x_{1:k}|\text{stop at }n) = \mathcal{N}\left(p_{x,\Sigma}(x_{1:k})\,\big|\, 0, \sigma(n; k)\right)\mathcal{N}\left(p_{y,\Sigma}(x_{1:k})\,\big|\, 0, \sigma(n; k)\right)
    \label{eq:stop_likelihood}
\end{align}

Using Eq.\eqref{eq:stop_likelihood} we can calculate target posteriors 
for an arbitrary maximum number of jets. This allows us to 
modify the likelihood loss by generalizing the Bernoulli splitting
variable $\delta_{in}$ to a continuous variable $y_i\in[0,1]$ and
override the troublesome $\psplit=0$ label for particle
$n_\text{max}$ by this estimate weighted by a hyperparameter $\lambda$,
\begin{align}
    \loss_\text{override} &= \left\langle -\sum_{i=1}^{n} \log \pkin(x_i|\nu_{i-1}) -\sum_{i=0}^n \lambda_i\log \pbin\left(y_i;  \rho_i\right)\right\rangle_{x\sim\pd} \\
    &\quad \text{with} \quad y_i(x_{1:n}) 
    = \begin{cases}
        (1-\delta_{in}) &i< n_\text{max}\\
        1-p_\text{fit}(\text{stop at }i|x_{1:i}) &i= n_\text{max}      
    \end{cases}\;,\\[4pt]
        &\quad \text{and} \quad \lambda_i= 1 - (1-\lambda)\delta_{in_\mathrm{max}}\;.    
\end{align}
The posterior $p_\text{fit}(\text{stop at }i|x_{1:i})$ is calculated
using Eqs.\eqref{eq:stop_posterior} and \eqref{eq:stop_likelihood} up to
8 jets. In practice, we find the best performance by
including a staircase scaling prior when calculating the posterior. To
match the dataset, we take $R_{(n+1)/n}=0.225$ and cap the
probabilities for $n<3$. Note that the hyperparameter $\lambda$ multiplies only the \pbin contribution for particle $n_\mathrm{max}$.

\begin{figure}[t]
    \includegraphics[width=0.325\textwidth, page=29]{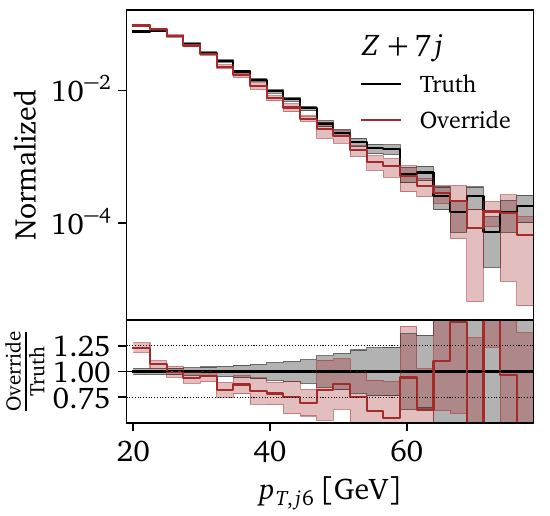} 
    \includegraphics[width=0.325\textwidth, page=33]{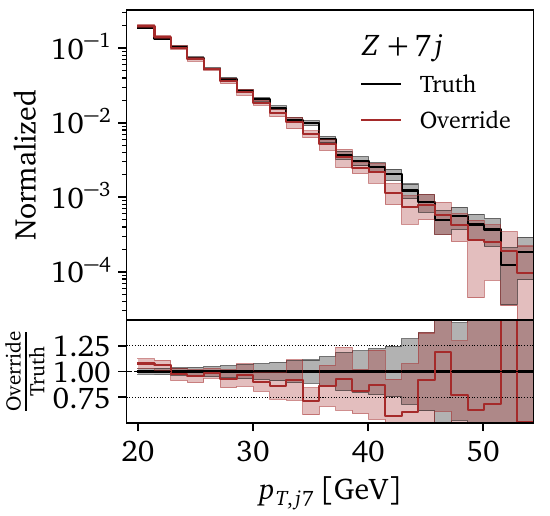} 
    \includegraphics[width=0.325\textwidth, page=4]{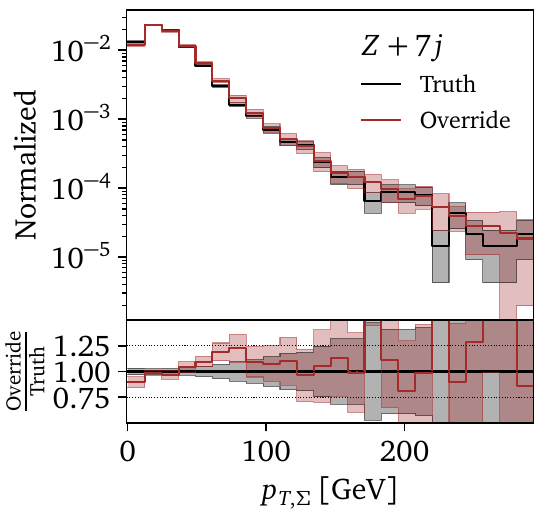} \\ 
    \includegraphics[width=0.325\textwidth, page=69]{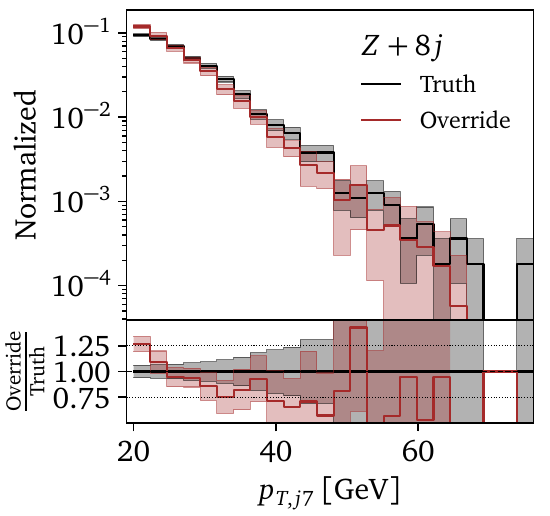}  
    \includegraphics[width=0.325\textwidth, page=73]{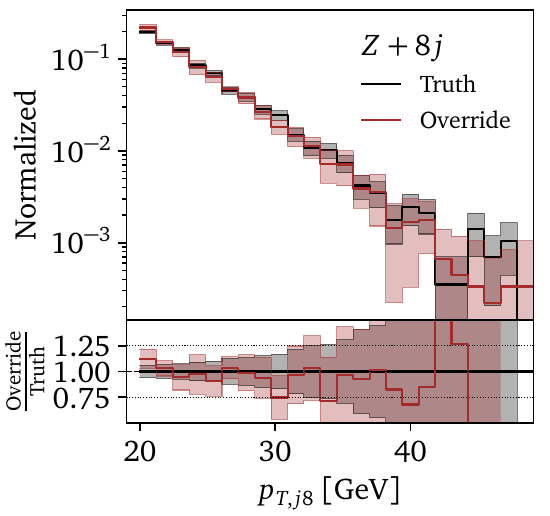}  
    \includegraphics[width=0.325\textwidth, page=8]{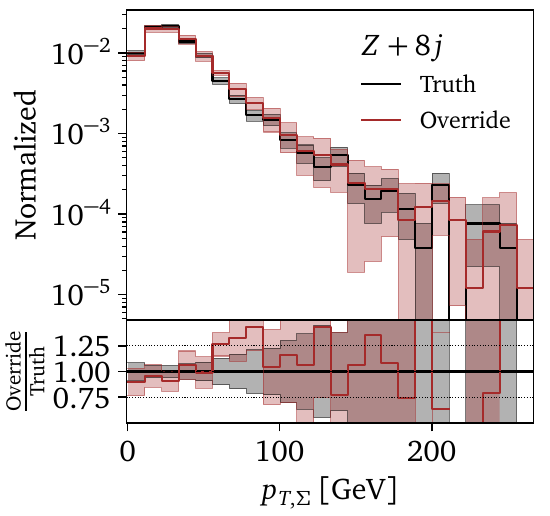} 
    \caption{Selection of features $Z + 7$ and 8-jet events, trained
      with $\loss_\text{override}$ on up to 6 jets.}
    \label{fig:samples_lambda_com}
\end{figure}

We train a network using the override loss with $\lambda=0.2$ and sample events in the same manner as before. In Fig.~\ref{fig:probs_stop_loss} (right), we show the event multiplicity
distribution. Similarly to the truncated loss, 
this override approach significantly 
increases the fraction of higher-multiplicity compared to the naive extrapolation. Looking at the 
kinematics of the network samples, shown  in
Fig.~\ref{fig:samples_lambda_com}, the $p_T$ distributions now display an excess toward low
values, but the global momentum correlation $p_{T,\Sigma}$ is reproduced to
greater accuracy compared to the case with the truncated loss. This is to be expected, since the override loss is specifically designed to match the global momentum distribution. Once again, this
demonstrates that autoregressive transformers can learn the universal
nature of jet radiation.

\subsubsection*{Code availability}

The code is available as part of the public Heidelberg hep-ml code and tutorial library \url{https://github.com/heidelberg-hepml/jetgpt-splittings}. The dataset is available upon request.

\section{Outlook}

The universality of splitting kernels and jet ratios in QCD
      provides the perfect physics question to see if appropriate
      generative networks can extrapolate.
      As an example, we study $Z+$jets events and the established staircase scaling of the jet number. The same procedure can be applied to generate the full set of jet constituents.

We employ an autoregressive transformer to learn a factorized
      likelihood for events across varying jet multiplicity. This
      autoregressive transformer sequentially predicts the kinematics of an
      additional jet along with the probability to radiate it. When trained the standard way, the transformer learns the
      kinematics of up to the 6 jets included in the training data
      with high fidelity. It also produces a small number of
      higher-multiplicity events with reasonable kinematics.

The first path towards extrapolation is to modify the
      training data with bootstrapping. This approach is
      straight-forward to adapt the multiplicity distribution, but
      some kinematic distributions are not learned very precisely.

Another way to extrapolate is to truncate the loss function and remove
      the explicit learning of the hard Sudakov factor for the highest
      multiplicity. Alternatively, we can override the hard Sudakov factor 
      using physics information.
      We find that both of these approaches are equally capable of
      generating high-quality events. These results establish that
      autoregressive transformers can learn the universal nature of
      jet radiation.

We emphasize that our study only shows that generative
      networks can extrapolate, given the right QCD properties of the
      training data. We expect the performance of all
      extrapolating networks to improve with technical
      advances. One idea for such an improvement might be the
      synchronous training of a transformer generator with a
      classifier, as described in Appendix~\ref{sec:discformer}.

\section*{Acknowledgements} 

We thank Nathanael Ediger and Maeve Madigan for their contributions to
earlier stages of this project, and Armand Rousselot and Sander Hummerich for valuable discussions on the DiscFormer.  AB and JMV are funded by the BMBF
Junior Group \textsl{Generative Precision Networks for Particle
  Physics} (DLR 01IS22079). JS\ is funded by the Carl-Zeiss-Stiftung through the project \textsl{Model-Based AI: Physical Models and Deep Learning for Imaging and Cancer Treatment}. The Heidelberg group is supported by the
Deutsche Forschungsgemeinschaft (DFG, German Research Foundation)
under grant 396021762 -- TRR~257 \textsl{Particle Physics
  Phenomenology after the Higgs Discovery}.  This work was also
supported by the DFG under Germany’s Excellence Strategy EXC 2181/1 -
390900948 \textsl{The Heidelberg STRUCTURES Excellence Cluster}.
Moreover, we would like to thank the Baden-W\"urttem\-berg Stiftung
for financing through the program \textsl{Internationale
  Spitzenforschung}, pro\-ject \textsl{Uncertainties – Teaching AI its
  Limits} (BWST\_ISF2020-010).  The authors acknowledge support by the
state of Baden-W\"urttemberg through bwHPC and the German Research
Foundation (DFG) through grant no INST 39/963-1 FUGG (bwForCluster
NEMO).

\clearpage
\appendix
\section{Improving likelihood training with dynamic reweighting}
\label{sec:discformer}

The traditional way to achieve ultimate precision in distributions obtained via generative networks has been via reweighting. However, this comes at the cost of having weighted events. These weights typically span orders of magnitude which, in turn, translate into small efficiencies $\epsilon = \langle w\rangle / w_\text{max}$ when performing accept-reject unweighting \cite{Backes:2020vka2}. These inefficiencies imply a reduction in statistical power, and thus reweighting might render the original motivation of speeding up event generation invalid, since now one might need to generate many more weighted events to obtain a similar number of unweighted events. One way to reduce event weights is to incorporate the discriminator information into the generator training. In the following sections we introduce and showcase the DiscFormer algorithm, inspired by the DiscFlow developed in Ref.~\cite{Butter:2021csz} for normalizing flows.

\subsection{Dynamic discriminator reweighting}
\label{sec:discformer_algorithm}

\begin{figure}[b]
    \centering
    \includegraphics[page=4,scale=0.48]{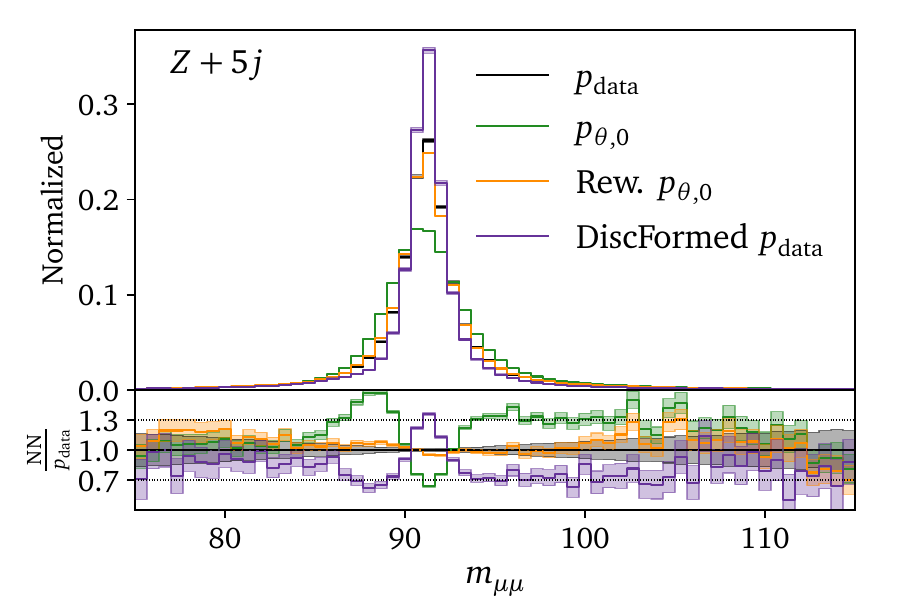}
    \includegraphics[page=63,scale=0.48]{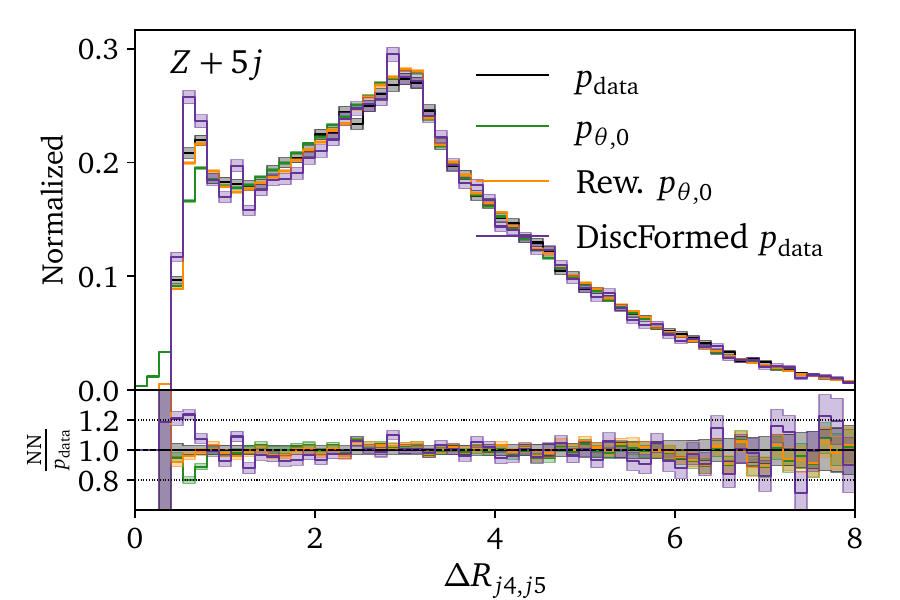}
    \caption{Distributions of the $Z$ boson mass (left) and $\Delta R_{j4,j5}$ (right) from the initial generator (green), the corresponding reweighted distribution $w_{0}\cdot \pmd(x)$ (orange) and the DiscFormed distribution $w_{0}\cdot \pd (x)$ (purple).}
    \label{fig:df_class_reweighting}
\end{figure}

The main idea behind the dynamic discriminator reweighting is to extend the usual log-likelihood loss with an event-dependent weighting factor $w(x)$
\begin{align}
    \mathcal{L} = \left\langle - w^{\alpha}(x) \log \pmd (x )\right\rangle_{x\sim \pd}.
    \label{eq:discformer_loss}
\end{align}
where $\alpha>0$ is a tunable hyperparameter and $\pmd(x)$ is the learned likelihood that depends on learnable network parameters $\theta$. The weighting factor $w(x)$ approximates the likelihood ratio $\pd(x)/\pmd(x)$ and is obtained from the score $D(x)$ of a neural discriminator. In practice, the weighting factor $w(x)$ in Eq.\eqref{eq:discformer_loss} amplifies the loss in regions where the generator is not precise enough. Intuitively, this can be thought of as modifying the training data distribution $\pd$ to increase the difference between $\pd(x)$ and $\pmd(x)$. This discriminator transformation, or DiscFormation, is visualized in Fig.~\ref{fig:df_class_reweighting}.

The neural discriminator $D(x)$ is trained to distinguish true samples, drawn from $\pd(x)$, from generated samples, drawn from $\pmd(x)$
\begin{align}
    \mathcal{L} &= \left\langle -\log D(x)\right\rangle_{x\sim\pd} + \left\langle -\log (1-D(x))\right\rangle_{x\sim\pmd} \notag\\
    &= -\int dx\pd (x) \log D(x) - \int dx\pmd (x)\log (1-D(x))\;.
\end{align}
We use variational calculus to find the minimum of this objective, yielding~\cite{Butter:2021csz,Rizvi:2023mws}
\begin{equation}
    0 = \frac{\delta\mathcal{L}}{\delta D(x)} = -\frac{\pd (x)}{D(x)} + \frac{\pmd (x)}{1-D(x)}\qquad \text{and}\qquad 
    w (x) = \frac{D(x)}{1-D(x)} = \frac{\pd(x)}{\pmd(x)}\;.
    \label{eq:df_classifier_weight}
\end{equation}
The assumption that $w(x)$ correctly approximates the likelihood ratio $\pd(x)/\pmd(x)$ can be validated by checking that the reweighted distributions correctly close onto a test set. We perform this test in Sec.~\ref{sec:classifier_reweight}, and plot the resulting reweighted distribution as the orange histogram in Fig.~\ref{fig:df_class_reweighting}. 

With this assumption, we can prove that the extended generator loss~\eqref{eq:discformer_loss} has the unique minimum $\pmd(x) = \pd(x)$. To this end, we insert the perfect discriminator criterion~\eqref{eq:df_classifier_weight} into the generator loss~\eqref{eq:discformer_loss} and add a Lagrange multiplier term to enforce the normalization of the learned likelihood $\pmd$, leading to the objective
\begin{align}
    \mathcal{L} = -\int{dx \ \pd(x) \ \left(\frac{D(x)}{1-D(x)}\right)^\alpha\ \log{\pmd(x)}} + \lambda \left(\int dx\ \pmd(x) - 1 \right).
    \label{eq:discflow_proof_loss}
\end{align}
In real-world implementations the learned likelihood $\pmd$ is constructed to satisfy the normalization constraint and the explicit Lagrange multiplier is not needed. For instance, we will construct $\pmd$ as a product of Gaussian mixture models which are normalized by construction. We now use variational calculus to find the minimum of this objective, yielding
\begin{align}
    0 = \frac{\delta\mathcal{L}}{\delta \pmd(x)} = -\left(\frac{D(x)}{1-D(x)}\right)^\alpha \frac{\pd(x)}{\pmd(x)} + \lambda = -\left(\frac{\pd(x)}{\pmd(x)}\right)^{\alpha +1} + \lambda\;.
    \label{eq:minimum_discformer_objective}
\end{align}
In the second equality we have again used the assumption of a perfectly trained discriminator, immediately yielding the unique solution $\pmd(x) = \pd(x)$, and $\lambda=1$.
We shall now discuss the behavior of this solution for several values of $\alpha$:
\begin{itemize}
\item The usual log-likelihood loss $\alpha = 0$ emerges as a smooth decoupling limit where the discriminator weight does not contribute to the loss.

\item In the limit $\alpha\to -1$, Eq.\eqref{eq:minimum_discformer_objective} becomes $-1+\lambda = 0$ and the optimization problem has no longer a unique solution. This is easy to understand, as the loss becomes 
\begin{align}
    \mathcal{L} = \left\langle -\log \pmd(x)\right\rangle_{x\sim \pmd},
\end{align}
where no information on $\pd(x)$ is included.
\item For $\alpha+1 <0$, the second derivative of the loss becomes negative. This means we can no longer train $\pmd(x)$ to approximate $\pd(x)$, since $\pmd(x) = \pd(x)$ is now a maximum of the loss:
\begin{align}
    \frac{\delta^2 \mathcal{L}}{\delta \pmd^2(x)} = (\alpha +1) \frac{1}{\pd(x)}\left(\frac{\pd(x)}{\pmd(x)}\right)^{\alpha +2} <0\quad \forall x.
\end{align}
\end{itemize}

In our experiments, we always set $\alpha = 1.0$ for the better training stability this choice offers, after checking that qualitatively the results behave in similar fashion for $0<\alpha\lesssim 5$.
In the subsections below, we describe in more detail how the DiscFormer training works, and demonstrate the capabilities of this approach in learning the full phase-space density of the $Z+5\text{ jets}$ dataset.

\subsection{Iterative DiscFormer algorithm}
\label{sec:discformer_iterative_implementation}
\begin{figure}
    \centering
    \begin{tikzpicture}[font=\small]
\usetikzlibrary{shapes.geometric, positioning, fit}

\node (gen_0) [rectangle, rounded corners, fill=Gcolor, minimum width=5cm, minimum height=2cm, align=center] {\textbf{Initial generator training}\\\\\normalsize{$\mathcal{L} = \left\langle -\log \pmd (x )\right\rangle_{x\sim \pd}$}};

\node (sample gen_i) [ellipse, rounded corners, fill=Gcolor, minimum width=6 cm, minimum height=2cm, align=center, right = 1.5cm of gen_0] {\textbf{Sample generator $\pmdj{i}(x)$}};

\node (dfc) [rectangle, rounded corners, fill=Bcolor, minimum width=6 cm, minimum height=2cm, align=center, below = 1.cm of sample gen_i] {\textbf{Discriminator training}\\\\\normalsize{$w_{i}(x) = \frac{\pd(x)}{\pmdj{i}(x)}$}};

\node (df) [rectangle, rounded corners, fill=Gcolor, minimum width=6 cm, minimum height=2cm, align=center, below = 1.cm of dfc] {\textbf{Re-train generator}\\\\\normalsize{{$\mathcal{L}^{\text{DF}}_{i} = \langle -w^\alpha(x)\log \pmd (x)\rangle_{x\sim \pd}$}}};

\draw [arrow, shorten >=.05cm] (gen_0) -- node [pos=0.5, above, anchor=south, yshift=0.cm] {$i\leftarrow 0$} (sample gen_i);
\draw [arrow] (sample gen_i) -- (dfc);
\draw [arrow] (dfc) -- (df);
\draw [arrow, shorten >=.05cm, shorten <=.05cm] 
  (df.east) -- ++(1.4, 0)
  -- ++(0, 6) 
  -- ($(sample gen_i.east) + (1.4, 0)$) 
  -- (sample gen_i.east) node[pos=0.4, above] {$i\leftarrow i+1$};

\node (df iteration text) [above = 0.1cm of sample gen_i, align=center] {$i^\mathrm{th}$ DiscFormer iteration};
\begin{scope}
  \node [fit={(sample gen_i)(dfc)(df)}, rectangle, rounded corners, draw=black, thick, inner sep=0.05cm] {};
\end{scope}

\end{tikzpicture}
    \caption{The iterative implementation of the DiscFormer algorithm. The re-training of the generator is done with the loss detailed in Eq.\eqref{eq:discformer_loss}.}
    \label{fig:discformer_algorithm}
\end{figure}
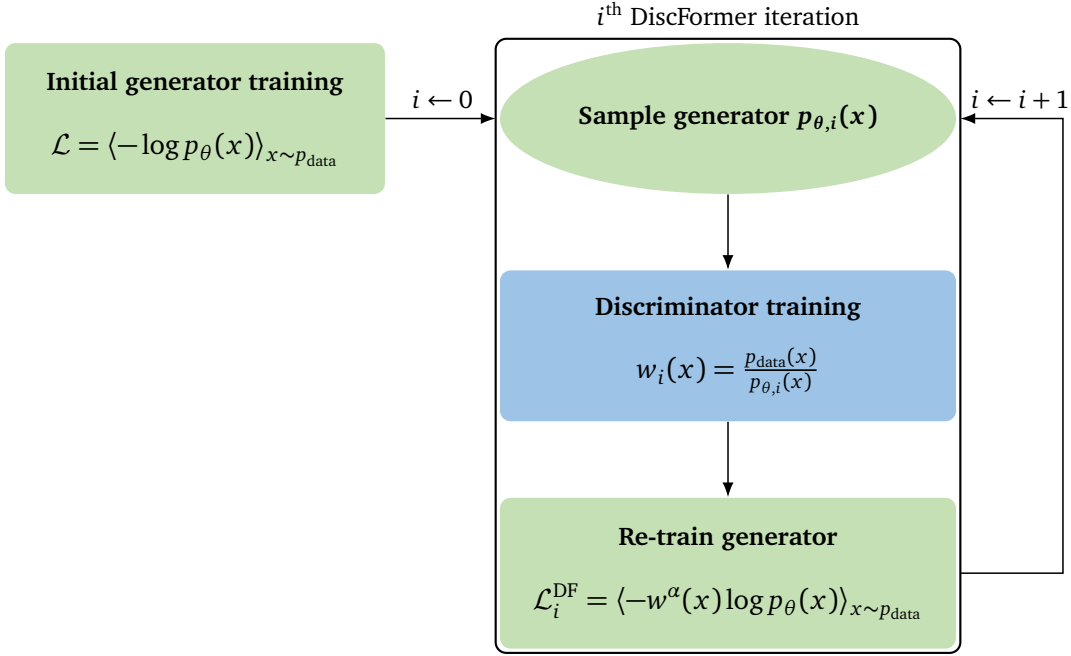

To train a generator with the DiscFormer approach described above, we require a discriminator $w(x) = \pd (x)/ \pmd (x)$ that tracks the state of the generator $\pmd$ throughout the generator training. Clearly, training such a discriminator for each generator update step is not a scalable approach. Ref.~\cite{Butter:2021csz} proposed a GAN-like approach where the generator and discriminator are trained jointly. Similar to GANs, this approach requires careful hyperparameter tuning to train the two networks jointly.

We propose an iterative procedure using discriminator checkpoints to avoid a joint training. To this end, we extract a checkpoint generator $\pmdj{0}(x)$ and use it to train a discriminator $D_0(x)$ to obtain the weights $w_0(x)$. Using the saved likelihoods of the checkpoint generator $\pmdj{0}(x)$ and the likelihoods of the current generator state $\pmd (x)$, we can reweight the checkpoint weights $w_0(x)$ to obtain the full weights $w(x)$
\begin{align}
w_0(x) &= \frac{\pd (x)}{\pmdj{0}(x)} = \frac{D_0(x)}{1-D_0(x)}\;\notag\\
w(x) &= \frac{\pd (x)}{\pmd (x)} = \frac{\pd (x)}{\pmdj{0}(x)}\frac{\pmdj{0}(x)}{\pmd (x)} = w_0 (x) \frac{\pmdj{0}(x)}{\pmd (x)}\;.
\end{align}
We emphasize that this checkpoint reweighting is exact, thanks to the fact that we can extract likelihoods from the generative model. The only approximation is in the assumption that the checkpoint discriminator $D_0(x)$ learns the likelihood ratio perfectly. Whenever the generator $\pmd (x)$ approximates the truth distribution $\pd (x)$ significantly better than the checkpoint generator $\pmdj{0}(x)$, it is important that we update the checkpoint generator to $\pmdj{1}(x)$ to avoid numerical issues from cancellations between $w_0(x)$ and $\pmdj{0}(x)/\pmd (x)$. This leads us to the loss function in the $i^\text{th}$ DiscFormer iteration
\begin{align}
    \label{eq:discformer_loss_iterative}
    \mathcal{L}^{\text{DF}}_{i} = \left\langle -w^\alpha(x)\log \pmd (x)\right\rangle_{x\sim\pd} = \left\langle -\left (\frac{D_{i}(x)}{1-D_{i}(x)}\frac{\pmdj{i}(x)}{\pmd (x)}\right )^{\alpha}\ \log \pmd(x)\right\rangle_{x\sim \pd}\;.
\end{align}

The implementation of the DiscFormer algorithm is depicted in Fig.~\ref{fig:discformer_algorithm}. We will describe now how the first iteration in the algorithm takes place. The process starts with an initial training of the generator, until its convergence is reached at state $\pmdj{0}(x)$. From this point, we start the DiscFormer iteration, represented as a black rectangle in Fig.~\ref{fig:discformer_algorithm}. We first draw samples from the generator, and train the discriminator to learn the likelihood ratio of $\pd(x)$ to $\pmdj{0}(x)$. To conclude the DiscFormer iteration, we re-train the generator using the DiscFormer loss, warm-starting it from $\pmdj{0}(x)$. This procedure can be repeated until convergence.

\subsection{Autoregressive transformer}
\label{sec:architecture_training}

The DiscFormer formalism developed above can be applied to any generative network that allows us to extract likelihoods $\pmd (x)$. Particularly attractive approaches are normalizing flows and parametric models like the generative transformer used in the main body of this work due to their ability of fast likelihood evaluation. We use an autoregressive transformer\cite{Butter:2023fov}, which is designed to interpret the phase space vector $x=(x_1,\ldots, x_{n})$ as a sequence of elements, and factorizes the $n$-dimensional probability into $n$ probabilities, successively conditioned
\begin{align}
\pmd(x) = \prod_{i=1}^n \pmd(x_i|x_1,...,x_{i-1})\; ,
\end{align}
The sequence elements are the kinematical quantities $(p_T,\phi,\eta,m)$, preprocessed as described in Sec.~\ref{sec:basics_data}. This autoregressive approach is mainly beneficial to our case because we can use our physics knowledge to group challenging phase space directions early in the sequence $x_1,...,x_n$. In contrast to the autoregressive transformer used in the main body of this work, this approach does not include splittings.

The network learns the factorizing conditional probabilities over phase space using a Gaussian mixture representation $\mathcal{G}$:
\begin{align}
  \pmd(x_i|\omega^{(i-1)} )= \sum_{G_{j}\in \mathcal{G}} w_j^{(i-1)} \normal (x_i; \mu_j^{(i-1)}, \sigma_j^{(i-1)} ), 
\label{eq:at_gaussians}
\end{align} 
where $\{w_j, \mu_j, \sigma_j\}$ are the components of the $j$-th gaussian.
In terms of architecture, our generator contains 80k parameters, consisting of 2 transformer decoder blocks, 4 self-attention heads and 50 Gaussian mixture elements. For more details, we refer the reader to Tab.~\ref{tab:df_hyperparameters}. The network is trained on approximately 80k and tested on 40k events from the $Z+5\text{ jets}$ dataset. We choose this working point to demonstrate the capabilities the DiscFormer approach can have when training data is scarce, compared to standard log-likelihood loss training. The challenges that the autoregressive generator faces for the $Z+5\text{ jets}$ dataset are: achieving percent level precision in the $Z$ mass peak and resolving the several hard $\Delta R$ boundaries between jets. For the latter, $5$ jets originate a total of $5(5-1)/2$ distinct $\Delta R$ features and the generator has to learn the subtle differences between them. The feature ordering in the sequence $x_1,...,x_n$ plays here a significant role in obtaining good performance for these features. We find that the best precision is achieved with
\begin{align}
    \{x_{i}\} = \{\phi, p_{T}, \eta, m\}_{j_5, j_4, j_3, j_2, j_1} \cup \{\phi, p_{T}, \eta\}_{\mu_{2}, \mu_{1}}.
\end{align}
We check that the $Z$ mass peak precision, on the other hand, is affected less by changes in the feature ordering. 

\subsection{Results}

\subsubsection*{Discriminator reweighting}
\label{sec:classifier_reweight}

In the derivation of the DiscFormer loss, we have made the assumption that $D(x)$ is a discriminator that correctly approximates the likelihood ratio of $\pd(x)/\pmd(x)$. We demonstrate in this subsection that the discriminator network is indeed capable of identifying failure modes of the generator $\pmd(x)$. Our discriminator is a transformer with 4 transformer-decoder blocks and 4 self-attention heads, summing up to about 800k parameters. We use the kinematic quantities $(p_T,\phi,\eta, m)$ for the final-state particles and the virtual $Z$ boson as input tokens, preprocessed in the same way as described in Sec.~\ref{sec:basics_data}. Additionally, we add tokens for the pairwise $\Delta R$ to inform the discriminator about this challenging correlation. Each time we train the discriminator, we do so with early stopping and a patience of 10 epochs. After each training, we check that the discriminator learns the correct likelihood ratio by reweighting the generated samples and checking that they close onto a test set. In Fig.~\ref{fig:df_class_reweighting} we show such closure test for the iteration 0 of one DiscFormer run. In this case, we find that the reweighted distribution $w_0(x)\cdot \pmdj{0}(x)$ correctly matches the truth.

\subsubsection*{DiscFormer}
\label{sec:discformer_reweight}

\begin{table}[tb]
    \centering
    \begin{tabular}[t]{l c c c c}
    \toprule
    Run \textbackslash ~AUC & $\pmdj{0}(x)$ & $\pmdj{3}^{\text{DF}}(x)$ & $\pmdj{3}(x)$ \\
    \midrule
    1 & 0.737 & 0.712 & 0.743 \\
    2 & 0.754 & 0.711 & 0.766 \\
    3 & 0.695 & 0.639 & 0.682 \\
    4 & 0.701 & 0.675 & 0.686 \\
    5 & 0.713 & 0.633 & 0.704 \\
    6 & 0.696 & 0.651 & 0.710 \\
    7 & 0.699 & 0.664 & 0.671 \\
    8 & 0.730 & 0.657 & 0.680 \\
    9 & 0.676 & 0.629 & 0.707 \\
    \midrule
    avg. $\pm$ std & $0.711\pm0.023$ & $0.663\pm0.029$ & $0.706\pm0.029$ \\
    \bottomrule
    \end{tabular}
    \caption{AUC values as quality metrics to study the DiscFormer performance, evaluated on 9 independent seeds. We
 show, from left to right, the initial generator $\pmdj{0}$, the final generator $\pmdj{3}^{\text{DF}}(x)$ after 3 discformer iterations, and the generator $\pmdj{3}(x)$ trained for the same number of epochs with a standard likelihood loss.}
    \label{tab:discformer_results}
\end{table}

Finally, we discuss the details of the DiscFormer experiment, where we modify the standard likelihood loss by incorporating discriminator information during training. The setup described below is designed such that we can compare the results from the DiscFormer approach directly to the standard likelihood loss training:
 \begin{enumerate}
 \item To start, we train the initial generator for 2500 epochs and extract the network with the best validation loss. We name this generator $\pmdj{0}(x)$.
 \item From the $\pmdj{0}(x)$ checkpoint, we compare two options: DiscFormer iterations, and standard training.
    \begin{itemize}
     \item DiscFormer trainings cover 3 iterations, each consisting of generator sampling, discriminator training and generator training with DiscFormer loss. The discriminator is trained from scratch at each iteration with early stopping and a patience of 10 epochs. The discriminator with best validation loss is also loaded for evaluation. The generator, on the other hand, is warm-started from the final state of the previous iteration and trained for 500 epochs. We call the final state of the generator at the end of the run $\pmdj{3}^{\text{DF}}(x)$.
     \item Standard generator trainings are continued from the same $\pmdj{0}(x)$ checkpoint for the same number of epochs as in the DiscFormer algorithm, i.e. $3\times 500 = 1500$. We name this generator $\pmdj{3}(x)$.
    \end{itemize}
 \end{enumerate}
 
 The final states of the generator $\pmdj{3}^{\text{DF}}(x)$ and $\pmdj{3}(x)$ are thus directly comparable, in the sense that they have been trained for the same number of epochs, and differences in their performance can be, in principle, attributed to the different training modalities. We find that the performance of the DiscFormer approach saturates after 3 iterations.

\begin{figure}[tb]
    \centering
    \includegraphics[page=4,scale=0.48]{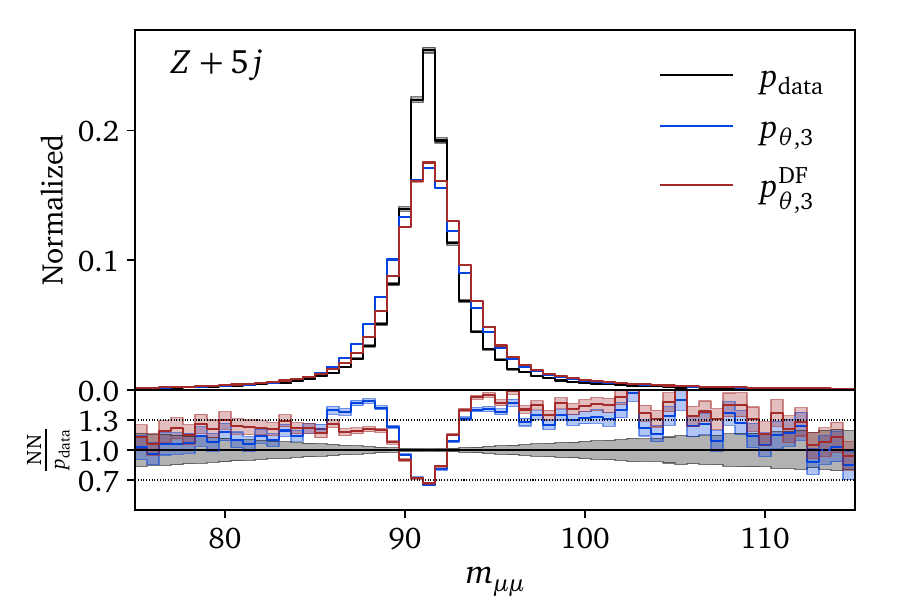}
    \includegraphics[page=63,scale=0.48]{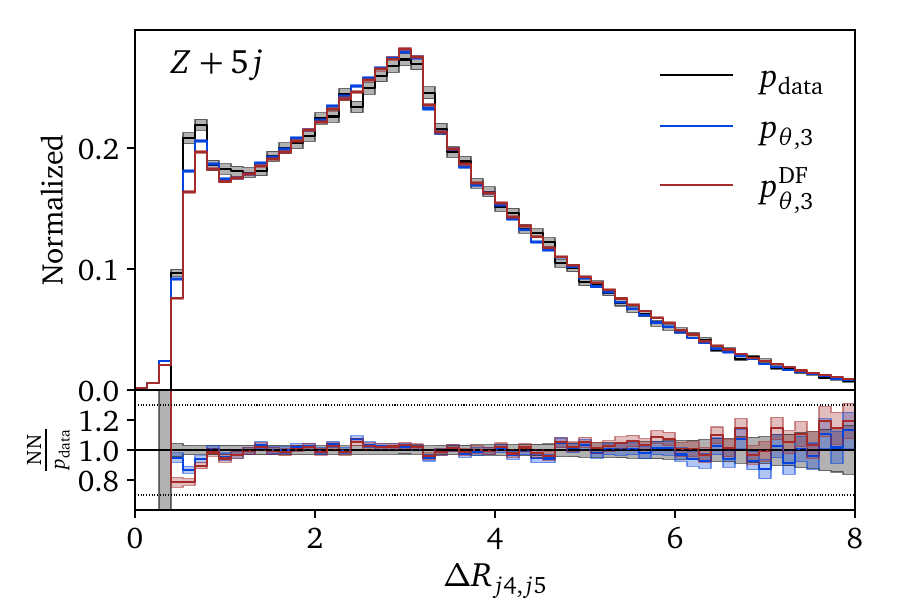}
    \caption{Distributions of the $Z$ boson mass (left) and $\Delta R_{j4,j5}$ (right) obtained from the standard generator training (blue) and from the generator after training for 3 DiscFormer iterations (red). Both networks have been trained for the same number of epochs.}
    \label{fig:discformer_results}
\end{figure}

 To systematically check whether performance is gained through the DiscFormer approach, we perform 9 identical runs where we compare the DiscFormer training versus the standard training. For these runs, we train discriminators with identical architecture to distinguish true data from samples generated from the corresponding $\pmd(x)$, and show the Area Under the Curve (AUC) of the Receiver-Operating Characteristic (ROC) curve obtained from evaluating these classifiers on a test set in Tab.~\ref{tab:discformer_results}. 
 We find that the discriminators trained to distinguish $\pmdj{3}^{\text{DF}}(x)$ from truth data have systematically lower AUC values than those trained to distinguish $\pmdj{3}(x)$ from truth. In particular, we observe that standard likelihood training improves the quality of the samples only marginally according to the discriminators, i.e. $\pmdj{0}(x)$ and $\pmdj{3}(x)$ showing very similar AUC values, whereas we observe a systematic improvement for the DiscFormer training $\pmdj{3}^{\text{DF}}(x)$.

\begin{figure}[b!]
    \centering
    \includegraphics[page=1,scale=0.65]{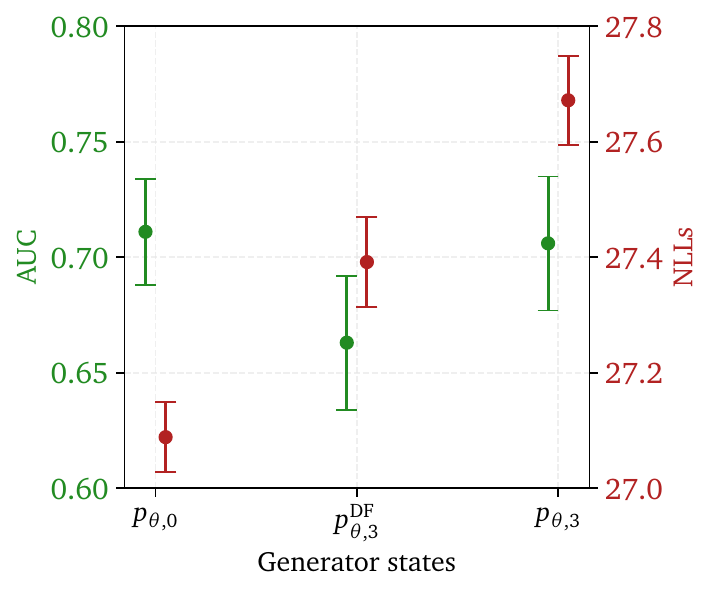}
    \caption{Mean and standard deviation of the AUC and negative log-likelihood (NLL) values evaluated on the same test set for the 9 runs from Tab.~\ref{tab:discformer_results}.}

    \label{fig:AUCs_NLLs}
\end{figure}

We can also check the performance of $\pmdj{3}(x)$ and $\pmdj{3}^{\text{DF}}(x)$ on two challenging phase space  features, the $Z$ mass and $\Delta R_{j4,j5}$, shown in Fig.~\ref{fig:discformer_results}. We observe that the generator trained with the DiscFormer approach performs slightly better in some of the regions near the peak of the $m_{Z}$ distribution, whereas the generator trained with the standard likelihood loss has become better in the tails. On the other hand, the $\Delta R_{j4,j5}$ sharp boundary seems to be slightly better resolved by the standard generator than by the generator trained with DiscFormer loss. More generally, we find no systematic performance differences in those phase space distributions.

As a third metric, we evaluate the negative log-likelihood (NLL) of all 3 generator stages on the same test set, and compute the mean and the standard deviation for the same 9 runs shown in Tab.~\ref{tab:discformer_results}. These, along with the mean and standard deviation of the AUC values, are shown in Fig.~\ref{fig:AUCs_NLLs}. We observe that the best NLL is generally obtained for the initial generator $\pmdj{0}(x)$. This makes sense, as this network was trained to minimize the NLL until the validation loss started to increase. In contrast, $\pmdj{3}(x)$ was trained beyond that point, leading to larger NLLs for the test data.
The DiscFormer $\pmdj{3}^{\text{DF}}(x)$ was trained as well beyond the minimum but with the target to minimize the DiscFormer loss. 
Nonetheless, the NLL is significantly better for the generator trained with the DiscFormer approach than for the generator trained with standard likelihood loss.

\section{Network hyperparameters}
\label{app:hyperparameters}

\begin{table}[ht!]
    \centering
    \begin{small} \begin{tabular}[t]{l l}
    \toprule
    Parameter & Value \\
    \midrule
    Optimizer & Adam\\
    Learning rate & $3 \cdot 10^{-4}$ \\
    LR schedule & constant \\
    Batch size & 512 \\
    \# Iterations & 200k \\
    \midrule
    \# Transformer Blocks & 3+3 \\
    Latent space size $d$  & 128\\
    \# Attention heads & 8\\
    \# Mixture model elements & 42 \\
    \# Trainable parameters & 1.2M \\
    \bottomrule
    \end{tabular} \end{small}
    \caption{Architecture and training hyperparameters. We use 3 blocks each for the particle-level transformer and the component-level transformer.}
    \label{tab:cfm_transfermer_hyperparams}
\end{table}

\begin{table}[ht!]
    \centering
    \begin{tabular}[t]{lcc}
    \toprule
    Parameter & Generator & Discriminator \\
    \midrule
    Optimizer & Adam & Adam \\
    Learning rate & $1 \cdot 10^{-4}$ & $3 \cdot 10^{-4}$ \\
    LR schedule (initial) & OneCycleLR & -- \\
    LR schedule (DiscFormer) & constant & ReduceLROnPlateau\\
    Batch size & 512 & 512 \\
    \# Iterations (initial) & until early stopping & -- \\
    \# Iterations (DiscFormer) & 5$\times$75k & until early stopping \\
    \midrule
    \# Transformer Blocks & 2 & 4 \\
    \# Attention heads & 4 & 4\\
    \# Mixture model elements & 50 & --\\
    \# Trainable parameters & 80k & 800k \\
    \bottomrule
    \end{tabular}
    \caption{Training hyperparameters and architecture for DiscFormer generator and discriminator.}
    \label{tab:df_hyperparameters}
\end{table}


\bibliography{tilman,refs_1,refs_2}

\end{document}